\documentclass[fleqn,usenatbib]{mnras}

\usepackage{newtxtext,newtxmath}

\usepackage[T1]{fontenc}

\DeclareRobustCommand{\VAN}[3]{#2}
\let\VANthebibliography\thebibliography
\def\thebibliography{\DeclareRobustCommand{\VAN}[3]{##3}\VANthebibliography}

\usepackage{graphicx}	
\usepackage{amsmath}	
\usepackage{adjustbox}
\usepackage{multirow}
\usepackage{tabularx}
\usepackage{booktabs}  

\usepackage{tikz}
\usetikzlibrary{shapes.geometric, arrows}

\usepackage{xcolor}  
\definecolor{blazeorange}{rgb}{1.0, 0.4, 0.0}
\definecolor{seagreen}{rgb}{0.18, 0.55, 0.34}
\definecolor{rufous}{rgb}{0.66, 0.11, 0.03}
\definecolor{royalfuchsia}{rgb}{0.79, 0.17, 0.57}
\definecolor{scarlet}{rgb}{1.0, 0.13, 0.0}
\definecolor{royalpurple}{rgb}{0.47, 0.32, 0.66}
\definecolor{darkblue}{rgb}{0, 0, 0.66}

\title[SSC Modeling of TeV Afterglows in GRBs]{Synchrotron Self-Compton Model of TeV Afterglows in Gamma-Ray Bursts}

\author[E. Aguilar-Ruiz et al.]{
Edilberto Aguilar-Ruiz,$^{1}$\thanks{E-mail: e.aguilar@irya.unam.mx}
Ramandeep Gill,$^{1,3}$\thanks{E-mail: r.gill@irya.unam.mx} 
Paz Beniamini,$^{2,3,4}$
and Jonathan Granot$^{2,3,4}$
\\
$^{1}$Instituto de Radioastronom\'ia y Astrof\'isica, Universidad Nacional Aut\'onoma de M\'exico, Antigua Carretera a P\'atzcuaro $\#$ 8701,  Ex-Hda. San Jos\'e de la \\ Huerta, Morelia, Michoac\'an, C.P. 58089, M\'exico\\
$^{2}$Department of Natural Sciences, The Open University of Israel, P.O Box 808, Ra'anana 4353701, Israel \\
$^{3}$Astrophysics Research Center of the Open university (ARCO), The Open University of Israel, P.O Box 808, Ra'anana 43537, Israel \\
$^{4}$Department of Physics, The George Washington University, 725 21st Street NW., Washington, DC 20052, USA
}

\date{Accepted XXX. Received YYY; in original form ZZZ}

\pubyear{\the\year{}}

\begin{document}
\label{firstpage}
\pagerange{\pageref{firstpage}--\pageref{lastpage}}
\maketitle

\begin{abstract}
The detection of a very-high-energy TeV spectral component in the afterglow emission of gamma-ray bursts (GRBs) has opened a new probe into the energetics of ultra-relativistic blast waves and the nature of the circumburst environment in which they propagate. The afterglow emission is well understood as the synchrotron radiation from the shock-accelerated electrons in the medium swept up by the blast wave. The same distribution of electrons also inverse-Compton upscatters the softer synchrotron photons to produce the synchrotron self-Compton (SSC) TeV emission. Accurate modeling of this component generally requires a computationally expensive numerical treatment, which makes it impractical when fitting to observations using Markov Chain Monte Carlo (MCMC) methods. Simpler analytical formalisms are often limited to broken power-law solutions and some predict an artificially high Compton-Y parameter. Here we present a semi-analytic framework for a spherical blast wave that accounts for adiabatic cooling and expansion, photon escape, and equal-arrival-time-surface integration, in addition to Klein-Nishina effects. Our treatment produces the broadband afterglow spectrum and its temporal evolution at par with results obtained from more sophisticated kinetic calculations. We fit our model to the afterglow observations of the TeV bright GRB\,190114C using MCMC, and find an energetic blast wave with kinetic energy $E_{k, \rm iso} = 9.1^{+7.41}_{-3.13} \times 10^{54} \, \rm erg$ propagating inside a radially stratified external medium with number density $n(r)\propto r^{-k}$ and $k=1.67^{+0.09}_{-0.10}$. A shallower external medium density profile ($k<2$) departs from the canonical approximation of a steady wind ($k=2$) from the progenitor star and may indicate a non-steady wind or a transition to an interstellar medium.
\end{abstract}




\section{Introduction}
The intense, short-lived, and highly variable \textit{prompt} $\gamma$-ray emission in gamma-ray bursts (GRBs) is followed by a much longer-lasting, temporally smooth, and spectrally broad band \textit{afterglow} \citep[see, e.g.,][for reviews]{Piran-99,Kumar-Zhang-15}. In the fireball model \citep{Goodman-86, Paczynski-86, Shemi-Piran-90}, the latter emission arises when the jet launched by the compact central engine interacts with the circumburst environment (CBM), i.e., interstellar medium (ISM) or wind-stellar medium, and produces an external forward shock \citep{Rees-Meszaros-92, Meszaros-Rees-93, Paczynski-Rhoads-93, Meszaros-Rees-97}. In the canonical afterglow scenario, as the shock-front sweeps up the CBM it accelerates electrons to relativistic Lorentz factors and into a power-law energy distribution, where they cool by emitting synchrotron radiation while gyrating in the shock-produced/amplified magnetic fields.

The afterglow synchrotron emission from a spherical outflow is now well understood \citep{Sari+98,Granot-Sari-02}. Its broadband spectrum (Radio-Optical-X-rays) is described by a multiply broken power law with smooth breaks at characteristic synchrotron self-absorption ($\nu_a$), injection ($\nu_m$), and cooling-break ($\nu_c$) frequencies. The same distribution of electrons producing the synchrotron emission is also expected to upscatter these seed synchrotron photons to high-energies ($\gtrsim0.1$\,GeV), and even very-high-energies ($\gtrsim0.1$\,TeV), via inverse-Compton \citep{Panaitescu-Meszaros-98,Chiang-Dermer-99,Dermer+00a,Dermer+00b,Panaitescu-Kumar-00}. The additional cooling affects the particle distribution and consequently the seed synchrotron spectrum \citep{Sari-Esin-01}, particularly when the scattering occurs in the Klein-Nishina regime \citep{Nakar+09}.

Long lasting (with observer-frame duration $T\lesssim35$\,ks) GeV emission has been detected by the \textit{Fermi}/LAT in several GRBs, and most notably in GRB\,130427A \citep{Ackermann+14}. This high-energy (HE) component initially overlaps with the prompt phase but shows smooth flux decline after the prompt emission ends, with $d\ln F_\nu/d \ln T\sim-1$, suggestive of it being the afterglow. The long sought after TeV emission was finally observed in GRB\,190114C by the Major Atmospheric Gamma Imaging Cherenkov (MAGIC) telescopes \citep{MAGIC_GRB190114C}. Detection of few more GRBs (160821B with a low detection significance of $\sim3\sigma$ \citep{Acciari+21}, 180720B \citep{Abdalla+19}, 190829A \citep{HESS_COLLABORATION+21}, 201216C \citep{Fukami+22}) at sub-TeV and TeV energies followed soon after by the MAGIC and high-energy Stereoscopic System (H.E.S.S) telescopes. Lastly, an unprecedented level of TeV emission was detected in the brightest GRB ever observed to date, GRB\,221009A, by the Large High Altitude Air Shower Observatory (LHAASO) that reported the detection of more than 64,000 VHE ($>0.2$\,TeV) photons within 3,000\,s of the burst trigger \citep{LHAASO+23}.

The TeV emission can most definitely be attributed to a separate spectral component, which is most likely synchrotron self-Compton (SSC) in leptonic models in which electrons are the main radiators \citep[see, e.g.,][for reviews]{Nava-21,Gill-Granot-22}. It cannot be synchrotron due to the maximum energy, $E_{\rm syn,\max}\simeq0.7(1+z)^{-1}\kappa\Gamma_1$\,GeV, to which synchrotron emission can be produced \citep{Guilbert+83,deJager-Harding-92}. Here $\kappa$ is the particle acceleration efficiency which is governed by the balance between shock acceleration and radiative cooling of particles at the shock front. As a result, the GeV emission can show these two overlapping components, i.e. synchrotron and SSC. In fact, GeV afterglow emission detected by the {\it Fermi}/LAT shows hardening of the spectral slope when compared to lower energies but still above the peak synchrotron energy. This is indicative of the overlap of the declining synchrotron spectrum and rising SSC spectrum, and can be seen most clearly in GRB\,190114C \citep{MAGIC_GRB190114C}.

The TeV spectrum of distant sources is affected by three distinct effects, one external and the other two internal to the emission region. The external effect includes the absorption of VHE photons by extragalactic background light (EBL) into producing $e^\pm$-pairs. Beyond a redshift of $z = 0.08$, the Universe starts to become opaque to TeV photons as they pair produce on the cosmic microwave background (CMB) and starlight UV photons. Internal to the emission region, the TeV photons can pair produce on softer photon, with energy $E_{\rm TeV}^{-1}\lesssim E \lesssim E_{\rm TeV}$, when the optical depth to $\gamma\gamma$-annihilation exceeds unity. This mostly affects super-TeV photons as the optical depth hardly ever exceeds unity for TeV photons at large distances from the central engine where the afterglow is produced. Finally, the VHE spectrum is affected by inefficient inverse-Compton scattering in the Klein-Nishina (KN) regime, when the energy of the incoming photon in the rest frame of the electron exceeds $m_ec^2/\gamma_e$, where $\gamma_e$ and $m_e$ are the electron Lorentz factor (LF) in the fluid frame and its rest mass, and $c$ is the speed of light. In this case, the VHE spectrum shows a spectral cutoff at that energy due to suppression in the scattering cross section. Most works account for this internal effect and the external one. 

The KN effect also alters the seed synchrotron spectrum \citep{Nakar+09,Jacovich1+21,McCarthy-Laskar-24}. In particular, it results in a hard particle distribution above the injection LF when particles are fast cooling, and consequently a hard spectrum above the injection frequency. Additional spectral break frequencies are also introduced. GRB afterglows are typically found in the slow-cooling regime several minutes after the initial prompt GRB. In this case, KN effects make the spectrum above the cooling break frequency harder, that results in obtaining shallower lightcurves, e.g., in the X-rays. These modifications to the spectrum and lightcurve affect the canonical closure relations (i.e. the expected spectral and temporal indices in a given power-law segment (PLS) of fast/slow cooling synchrotron emission) and yield inconsistent results when KN effects are not accounted for. For example, outflow kinetic energy estimates obtained from X-rays and GeV afterglows may differ when X-ray emitting electrons are IC cooled and IC cooling for GeV emitting electrons is KN suppressed \citep{Beniamini2015}. 

To make model comparisons with afterglow observations, it is advantageous to have a fast model-fitting code. This often requires an analytic formalism that comes with several assumptions to make the computation faster. When fitting a synchrotron plus SSC model to observations, the most common assumptions include considering only the emission along the line-of-sight (LOS), sharply broken spectrum with asymptotic spectral slopes, a step-function treatment of the Compton scattering cross-section, and simpler radiative transfer with instantaneous escape of radiation and no adiabatic cooling of particles.

Several efforts have been made using some/all of these assumptions to provide a semi-analytic prescription for the calculation of an SSC spectrum when including KN effects \citep[e.g.,][]{Nakar+09, Jacovich1+21, Yamasaki-Piran-22, McCarthy-Laskar-24, Pellouin+24}. All of these approaches are inherently limited in accuracy due to the simplifying assumptions that they make. In particular, as shown in the current work, they all yield a high Compton-Y parameter that tends to enhance the energy radiated in the SSC component. 
This leads to inferring lower values of the ratio $\epsilon_e/\epsilon_B$, a lower cooling LF, $\gamma_c$, and therefore, a lower $\nu F_\nu $ peak frequency of the slow-cooling synchrotron spectrum. In fast cooling, the reduction of $\nu_c$ produces a more extended low-frequency spectral tail below $\nu_m$. When the assumption of power-law spectrum with sharp breaks is made, the model is incapable of describing the smooth spectral shape near the break frequencies that in some spectral regimes are clustered near each other \citep[see, e.g.,][]{Yamasaki-Piran-22}. Typically, such works also only include emission along the LOS, an approximation that may be valid for ultra-relativistic spherical flows for which the emission comes from within a narrow beaming cone of angular size $1/\Gamma$ around the LOS. However, this simplification breaks down in jets with angular structure where the angular size of the beaming cone may vary with polar angle due to the angular dependence of the bulk LF, $\Gamma(\theta)$. 

An alternative approach that overcomes these limitations and improves the accuracy, while avoiding an explicit calculation of the Compton-Y parameter, is by numerically solving the coupled kinetic equations describing the evolution of particles and photons in both energy and time \citep[e.g.][]{Petropoulou-Mastichiadis-09,Fukushima+17, Derishev-Piran-21, Hope+25}. In this framework, the full KN cross-section is used and the KN effect on the particle distribution and the corresponding modification to the synchrotron and SSC spectrum is self-consistently accounted for. In addition, this approach allows for the inclusion of adiabatic cooling of particles and other non-linear processes that are often challenging to treat analytically, namely $e^\pm$-pair production via $\gamma\gamma$-annihilation and synchtron self-absorption. The numerical treatment also allows the integration over the equal-arrival-time-surface (EATS). Generally, such methods are computationally intensive and are not suitable for spectral and/or lightcurve fitting to observations.

The two approaches do not always yield consistent results. For example, \cite{Hope+25} showed that the SSC cooling rate obtained with their kinetic approach deviates from the results obtained with the analytic approach of \cite{McCarthy-Laskar-24}, particularly at early times and low electron energies. Similarly, \cite{ Derishev-Piran-21} compared their kinetic model results with other approaches finding differences in the predicted spectrum. Therefore, while analytical approaches offer vastly superior computational performance, making them ideal for parameter estimation using MCMC type analyses, they lack in accuracy when compared with results obtained from kinetic approaches.

In this work, we propose a semi-analytic approach that addresses the several issues faced by analytical treatments and yields more accurate time-dependent synchrotron and SSC spectra with only a modest computational cost and time investment. Our model solves both photon and electron continuity equations analytically, assuming a quasi-steady state over timescales much shorter than the smallest time over which both distributions are modified. Our model incorporates the effects of adiabatic expansion and photon escape. Both are often not included in analytic treatments and modify the broadband spectrum in important ways. We numerically calculate the Compton-Y parameter adopting the exact KN cross-section as well as the SSC spectral radiated power. Finally, the observed spectrum is obtained by performing integration over the EATS. 

The structure of the paper is as follows. Section \ref{sec::THE MODEL} describes the physical model and the underlying assumptions to construct a semi-analytic model. In Section \ref{sec::MODEL VALIDATION}, we first validate our model by comparing the synchrotron-only emission with the standard analytic models. Then, we compare our model with a kinetic approach over a wide dynamical range in time. Further, we demonstrate that ignoring photon escape and adiabatic escape impacts significantly on the resulting SSC spectra . Finally, we compare our model with the standard analytic model for SSC with KN effects and we discuss the discrepancies. Section \ref{sec::MCMC_FITS} presents the results of fitting the observed emission of the GRB 190114C. Finally, in Section \ref{sec::CONCLUSIONS}, we discuss our results and summarize our conclusions.

\section{The Model}\label{sec::THE MODEL}
\subsection{Forward Shock Dynamics in the Thin-Shell Approximation}
We consider the dynamical evolution of an ultrarelativistic, thin spherical shell propagating through an external medium with radially stratified number density
\begin{equation}
    n(r) = n_0 \left( \frac{r}{r_0}\right)^{-k} \, ,
\end{equation}
where $n_0$ is the normalization at $r_0 = 10^{18}$\,cm when the index $0 < k < 2$. The case $k=0$ ($k=2$) implies a constant interstellar medium (wind) profile, expected to be valid for short/hard (long/soft) GRBs. The mass density of the external medium is then expressed as $\rho (r) = A r^{-k}$, with $A= n_0 r_0^k m_p$, where $m_p$ is the proton mass. Initially, the shell is coasting at a bulk LF $\Gamma_0\gg1$ and starts to sweep up the external medium with mass $m(r) =[4\pi/(3-k)]\rho(r) r^3$ and accelerate it to a proper velocity $u\approx u_0 = \Gamma_0\beta_0$, where $\beta_0 = (1 - \Gamma_0^{-2})^{1/2}$. The interaction of the shell with the external medium produces a double shock structure, where a \textit{forward} shock propagates ahead of the shell and shock heats the swept up material, and a \textit{reverse} shock propagates through the ejecta shell and slows it down while extracting its kinetic energy. During the coasting phase the kinetic energy in the system resides in both the ejecta and swept up material, such that $E_{k, \rm iso} = (\Gamma_0-1) M_0 c^2 = [\Gamma(r)-1]M_0c^2 + u(r)^2m(r)c^2$, where $M_0$ is the baryonic load of the ejecta.
Once the shocked swept up mass reaches $m(r) \approx M_0/\Gamma_0$, the shell of ejecta and swept up material must decelerate at the deceleration radius,
\begin{equation} \label{eq_rdec}
    r_{\rm dec} = \left[ \frac{(3-k) E_{k,\rm iso}}{4\pi \, A \, c^2 u_0^2}  \right]^{1/(3-k)}\,,
\end{equation}
where most of the kinetic energy of the ejecta is transferred to that of the shocked swept up material. 
At $r > r_{\rm dec}$, the shell follows the self-similar \citet{Blandford-McKee-76} solution and its complete 
dynamical evolution, from the coasting to the non-relativistic expansion phase, can be obtained 
from energy conservation to yield \citep[e.g.,][]{Panaitescu-Kumar-00,Gill-Granot-18b}
\begin{equation}
    \Gamma(\xi) = \frac{\Gamma_0+1}{2} \xi^{k-3} \left[ 
    \sqrt{1 + \frac{4\Gamma_0}{\Gamma_0 +1 } \xi^{3-k} + \left( \frac{2\xi^{3-k}}{\Gamma_0+1} \right)^2 } - 1\right] \, ,
\end{equation}
with $\xi = r/r_{\rm dec}$ is the normalized radius.
The shell becomes non-relativistic when $\xi > \xi_{\rm nr}$, where $\xi_{nr} = (\Gamma_0^2/3)^{1/(3-k)}$ establishes the beginning of the non-relativistic regime with 
$\Gamma(\xi_{nr}) = 2$ \citep{Panaitescu-Kumar-00}.

When using the thin-shell approximation, we make the simplifying and explicit assumption that the bulk Lorentz factor of the fluid behind the shock is the same as that of the shock front\footnote{In reality, the ultrarelativistic shock moves slightly faster with $\Gamma_{\rm sh} = \sqrt{2} \Gamma$, and when making this distinction $\Gamma_{\rm sh}$ should be used to calculate the shock radius for a given lab-frame time.}, i.e. $\Gamma_{\rm sh} = \Gamma$. In this approximation, the radial width of the shock-front in the comoving frame (all comoving quantities are henceforth primed) can be obtained from particle number conservation. This yields
\begin{equation}
    \Delta ' (r) = \frac{r}{4(3-k) \Gamma} \, .
\end{equation}
The total internal energy density of the shocked fluid just behind the shock is given by $u_{\rm int}'(r) = \left(\Gamma-1 \right)\, n' \, m_p c^2$, where the number density of the shocked fluid, $n'$, can be related to that of the external medium by the shock jump condition, so that for a strong shock 
\begin{equation}
    n'(r) = \frac{ \hat{\gamma}_{\rm ad} \Gamma + 1 }{\hat{\gamma}_{\rm ad} - 1} \, n  = 4 \, \Gamma \, n \, .
\end{equation}
when the adiabatic index is described as $\hat{\gamma}_{\rm ad} = (4\Gamma + 1)/3\Gamma$ \citep[e.g.][]{Kumar-Granot-03}, which equals $4/3$ ($5/3$) for a relativistic (non-relativistic) gas.
\subsection{Non-Thermal Electrons}

\subsubsection{Electron injection}
The dominant radiation mechanism for the broadband afterglow emission in GRBs is synchrotron emission from non-thermal, relativistic electrons. These are accelerated into a power-law energy distribution at the shock front, and they are injected into the emission region at the following rate, 
\begin{equation}\label{eq_e_InjectionDistribution}
\frac{dn_{e,\rm inj}'(\gamma)}{dt'}\equiv{q_e'}(\gamma, r) = {q_0'} (r) \,
\gamma^{-p} \, 
\qquad \gamma_{m}(r) \leq \gamma \leq \gamma_{M}(r) \, , 
\end{equation}
where $q_0'$ is the injection rate normalization, and $\gamma_ {m}$ and $\gamma_{M}$ are the minimum and maximum electron LFs, respectively. By assuming that a fraction $\epsilon_e$ of the total internal energy density of the shocked fluid goes into accelerating electrons and taking $p>2$, the LF of the minimal energy electrons can be obtained \citep{Sari+98},
\begin{equation}\label{eq_eLF_min}
    \gamma_{m}  = \frac{(p-2) \, m_p \, }{(p-1) \, m_e \,  } \, \frac{\epsilon_e}{ \xi_e }( \Gamma - 1 ) \, ,     
\end{equation}
where $m_e$ is the electron mass and $\xi_e$ represents the fraction of the total number of shock heated electrons that are accelerated into a power-law energy distribution. The maximum Lorentz factor $\gamma_M$ is obtained from the equilibrium between radiative cooling, including SSC, and shock acceleration, as discussed in Eq.\,\ref{eq_eLF_MAX} below.

\subsubsection{Electron cooling}

The relativistic collisionless forward shock amplifies any pre-existing magnetic field in the external medium, and more importantly, generates new magnetic field via current-driven instabilities. In the standard afterglow model, it is generally assumed that a fraction $\epsilon_B$ of the internal energy density of the shocked fluid goes into that of the magnetic field, with $\epsilon_B u_{\rm int}' = B'^2/8\pi$, so that the strength of the comoving magnetic field is
\begin{equation}
    B'(r)  = (32\pi m_p c^2 )^{1/2} \, {n}^{1/2} \, \epsilon_B^{1/2} [\Gamma(\Gamma-1)]^{1/2} \, .  
\end{equation}
As the shock-accelerated relativistic electrons gyrate around the magnetic field lines, they emit synchrotron radiation that cools the electrons at the rate 
\begin{equation}
    \dot{\gamma}_{\rm syn}(\gamma) = -\frac{\gamma^2}{t'_{\rm  B}(r)} \, , \quad
    \text{with}
    \quad
    t'_{\rm  B} = \frac{6\pi m_e \, c}{\sigma_{\rm T} \, B'^2}  \, .
\end{equation}
If synchrotron radiation is the dominant coolant of electrons, then the LF of electrons that are cooling at the dynamical time,
\begin{equation}
    t_{\rm dyn}'(r) = \int_0^r \frac{dr'}{ \Gamma(r') \beta(r') c }\,,
\end{equation}
is given by $\gamma_{c}^{\rm syn} = t'_{\rm B} / t_{\rm dyn}'$. If the same distribution of electrons also inverse-Compton scatters the produced synchrotron radiation, the overall cooling rate is enhanced by the Compton-Y parameter, $Y_{\rm ssc}(\gamma)$, so that 
\begin{equation} \label{eq_e_loss_rate_SSC}
    \dot{\gamma}'_{\rm ssc}(\gamma) = -\frac{\gamma^2}{t'_B(r)} \,  \left[ 1 + Y_{\rm ssc} (\gamma) \right] \, ,
\end{equation}
where \citep[e.g.,][]{Nakar+09}
\begin{align}   \label{eq_Yssc_PARAM_DEF}
    Y_{\rm ssc}(\gamma) 
    &\equiv \frac{P_{\rm IC}(\gamma)}{P_{\rm syn}(\gamma)} \,,
\end{align}
and $P_{\rm syn}$ and $P_{\rm IC}$ are the radiated powers in synchrotron and IC emissions. Finally, as the shell expands, particles are further cooled adiabatically by doing $PdV$ work on the shell, for which the cooling rate is \citep[e.g.,][]{Dermer-Humi-2001, Longair-2011, Hope+25}
\begin{equation} \label{eq_e_loss_rate_ADIBATIC}
    \dot{\gamma}'_{\rm ad}(\gamma) = 
    -\frac{(\hat{\gamma}_{\rm ad} -1 ) }{t'_{\rm ad} (r)}  \, {\gamma} \, ,
    \quad \text{with} \quad
    \frac{1}{t'_{\rm ad }} =   \frac{3\Gamma \beta \, c}{r} \left( 1 - \frac{1}{3}\frac{d\log\Gamma}{d\log r} \right) \,,
\end{equation}
where
\begin{equation}\label{eq_dGamma_dR}
    \frac{d\log\Gamma}{d\log r} = \frac{d\log\Gamma}{d\log\xi} = (k-3)
    \left[ 
    1 - \frac{ 2 + \Gamma_0\left({\Gamma_0+1}\right) \xi^{k-3} }
    {2 \Gamma(\xi) \left\{ \Gamma(\xi) +  (\Gamma_0 +1 ) \xi^{k-3} \right\}}  \right] \, .
\end{equation}
Due to the additional cooling by the synchrotron radiation field, the electron cooling LF is determined from Eq.\,\ref{eq_e_loss_rate_SSC} and given by the condition 
\citep[e.g.][neglecting adiabatic losses]{Nakar+09}
\begin{equation} \label{eq_eLF_ssc}
    \gamma_c^{\rm ssc} \left[ 1 +  Y_{\rm ssc} (\gamma_c^{\rm ssc}) \right] = \gamma_{c}^{\rm syn} \,.
\end{equation}
The use of the above expression to obtain $\gamma_c$ is only justified in the thin-shell approximation, where it is assumed that particles cool at the same location, in a very 
thin layer behind the shock, where they are accelerated. In fact, this approximation is only truly valid for fast-cooling particles. During most of the afterglow phase, particles 
are slow-cooling and the emission is produced over a much larger volume behind the shock, and $\gamma_c$ evolves due to the evolution in the magnetic field energy density downstream of the shock. To account for this possibility, some works instead use an approximate relation $t_{\rm dyn}' = t_{\rm lab}(r)/\Gamma(r)$ \citep[e.g.][]{DeColle+12a,Gill-Granot-18b}, where $t_{\rm lab}(r) = \int_0^r dr'/\beta(r')c$ is the lab frame time corresponding to the distance traveled by the thin shell.

\subsubsection{Electron distribution }
Once electrons are injected inside the radiative zone, their evolution is governed by the continuity equation, 
\begin{equation}
     \frac{\partial n_e'(\gamma,t')}{\partial t'}  + 
     \frac{\partial}{\partial\gamma} \left[  \dot{\gamma}'_{\rm cool}(\gamma)\, n_e'(\gamma,t')\right] 
     = q_e'(\gamma, t')  - \frac{n_e'(\gamma, t')}{t'_{\rm ad}(t')}\,,
\end{equation}
where the last term on the R.H.S is for adiabatic density dilution. The total electron cooling rate is $\dot{\gamma}_{\rm cool}' = \dot{\gamma}_{\rm ssc}' + \dot{\gamma}_{\rm ad}'$, due to SSC and adiabatic expansion.

%
%
The continuity equation is a non-linear equation and its solution generally requires a numerical treatment. 
Here we simplify it by assuming that for $\Delta t'\ll t_{\rm cool}'$, where $t_{\rm cool}'$ is the shortest 
timescale over which the particle distribution is modified, the continuity equation can be approximately 
treated using a quasi-steady state approximation, i.e. $\partial n_e'(\gamma,t')/\partial t' \approx 0$. 
In this case, it admits the following solution at a given radius (see appendix \ref{Appendix:cont_eq_sol} for details) 
\begin{equation}\label{eq_eD_steady_state}
    n_{\rm e}'(\gamma) =\frac{1}{|\dot{\gamma}_{\rm cool}  (\gamma)|} 
    \begin{cases}
    \int_{\gamma_m}^{\gamma_{\rm M}}I_{\rm e}(\gamma') \, q'_{\rm e} (\gamma') d\gamma' \,, &  \gamma_{m'} < \gamma < \gamma_m
    \\
    \int_{\gamma}^{\gamma_{\rm M}}I_{\rm e}(\gamma') \, q'_{\rm e} (\gamma') d\gamma'\,, & \gamma_m < \gamma < \gamma_M \, ,    
    \end{cases}
\end{equation}
where the integrating factor is
\begin{equation}\label{eq_Ie}
    I_{\rm e}(\gamma') = 
    \exp\left[{  -\frac{1}{t'_{\rm ad}} \int_{\gamma}^{\gamma'} \frac{1}{ \, |\dot{\gamma}'_{\rm cool}    (\gamma'')| } d\gamma'' }\right] \, .
\end{equation}
The solution in the top row gives the particle density of those particles that have already cooled below $\gamma_m(r)$ to $\gamma_{m'}(r)$, and the bottom row yields the solution for all the particles that have cooled to $\gamma\geq\gamma_m(r)$.

As electrons are injected above $\gamma_m$, they cool and move to lower energies over a dynamical time due to adiabatic and SSC cooling. We can estimate the minimal particle LF, $\gamma_{m'}(t')$, by equating the step time, $\Delta t'$, with the time required for electrons to lose energy from $\gamma_{m'}(t'-\Delta t')$ to $\gamma_{m'} (t')$, 
\begin{equation}
    \Delta t' = -\int_{\gamma_{m'}(t'-\Delta t' )}^{\gamma_{ m'}(t')} \frac{d\gamma}{ \, |\dot{\gamma}_{\rm cool} (\gamma)| }\,.
\end{equation}

The solution in Eq.\,\ref{eq_eD_steady_state} is general and describes the so-called \textit{slow} and \textit{fast} cooling regimes including adiabatic expansion. When cooling is dominated by SSC, the cooling timescale is very small compared with the adiabatic timescale. In that case, the factor $I_e \sim 1$ and we obtain the well known expression for the fast-cooling regime \citep[e.g.][]{Nakar+09}, with an additional component below the SSC cooling particles, 
\begin{equation}\label{eq_eD_fast}
    n_e'(\gamma) \propto 
        \begin{cases}
           \gamma^{2}, & \gamma \ll \gamma_c^{\rm ssc} \\
           \frac{1}{1+Y_{\rm ssc}(\gamma)}
    \gamma^{-2}, & \gamma_c^{\rm ssc} \ll \gamma < \gamma_m \\
           \frac{1}{1+Y_{\rm ssc}(\gamma)}
     \gamma^{-p-1} , & \gamma > \gamma_m
        \end{cases}
\end{equation}
where the strict power laws are valid in the asymptotic regime, i.e. away from any breaks in the distribution.
Similarly, when adiabatic cooling is the dominant cooling mechanism up to $\gamma_c^{\rm ssc}$, with $\gamma_c^{\rm ssc} \gg \gamma_m$, then 
$I_e \propto \left({\gamma/\gamma'}\right)^{3}$, and the solution is 
\begin{equation}\label{eq_eD_slow}
    n_e'(\gamma) \propto
    \begin{cases}
        \gamma^{2}, & \gamma < \gamma_m \\
        \gamma^{-p}, & \gamma_m < \gamma \ll \gamma_c^{\rm ssc}\\
        \frac{1}{1+Y_{\rm ssc}(\gamma)} \gamma^{-p-1} & \gamma \gg \gamma_c^{\rm ssc}
    \end{cases},
\end{equation}
which is the power-law behavior  expected in the slow-cooling regime with an additional component below $\gamma_m$.
%
%
The normalization for the electron distribution is obtained from the shock jump condition, yielding
\begin{equation}\label{eq_eD_normalization}
    4\Gamma(r)n(r) =
\int_{1}^{\gamma_M} d\gamma \, n_e'(\gamma,r) \frac{}{} \, .
\end{equation}

The maximum LF up to which electrons can be accelerated is determined by comparing the shock acceleration timescale with that for radiative losses, i.e., $t_{\rm acc}' = t_{\rm cool}'$, with $t_{\rm acc}' \simeq \frac{\gamma m_e c}{\eta_{\rm acc} e B'}$, 
where $\eta_{\rm acc}$ is the acceleration efficiency, such that 
\begin{equation}\label{eq_eLF_MAX}
    \gamma_M \left[ 1 +  Y_{\rm ssc} (\gamma_M) \right]^{1/2} = \left( \frac{6\pi e \, \eta_{\rm acc} }{\sigma_{\rm T} B'} \right) ^{1/2}\, ,
\end{equation}
where $e$ is the elementary charge. 
The Compton-Y parameter declines at high electron LFs due to the Klein-Nishina suppression, and therefore at large 
$\gamma$ 
the maximum electron energy can be simply calculated by taking $Y_{\rm ssc}(\gamma_M) = 0$.

%
%
\subsection{The Compton-Y Parameter}\label{sec::Compton_parameter}

\begin{figure}
	\includegraphics[width=\columnwidth]{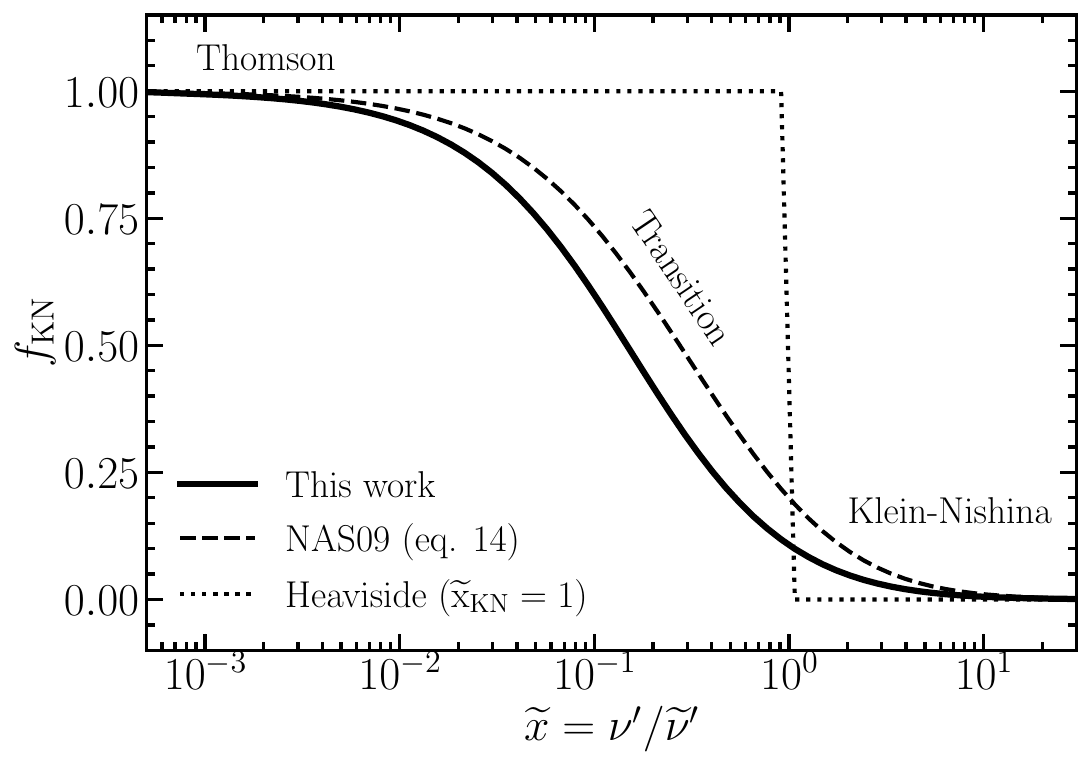}
    \vspace{-16pt} 
    \caption{Klein-Nishina scattering kernel for different approximations. The solid line is the expression provided in Eq. \ref{eq_fKN_this_work} and used in this work. The dashed line is the approximation from Eq.\, 14 in \citet{Nakar+09}. The dotted line is the Heaviside approximation with $\widetilde{x}_{\rm KN} = 1$ (see Eq. \ref{eq_fKN_step_approx}), which is very crude and employed in most analytical works.
    }
    \label{fig:fKN}
\end{figure}

The shape of the electron distribution and therefore the shape of the SSC spectrum are strongly modified by the value of the Compton-Y parameter when $Y_{\rm ssc}(\gamma) > 1$. The Compton-Y parameter is defined in Eq.\ref{eq_Yssc_PARAM_DEF} and can be expressed more generally for an isotropic radiation field in the fluid frame interacting with isotropic and relativistic electrons, (see appendix \ref{Appendix:Compton_Param_derivation} for more details)
\begin{equation}
    \label{eq_Yssc_PARAM}
    Y_{\rm ssc}(\gamma, r) = 
    \frac{1}{u_{B}' (r)}    
    \int_{0}^{ \infty } d\nu' \, u_{\nu'}'^{\rm syn}(\nu', r) 
    \, f_{\rm KN}\left( \frac{\nu'}{\widetilde{\nu}'}\right) \, ,
\end{equation}
where $u_B' = B'^2/8\pi$ is the magnetic field energy density in the shell, $u_{\rm \nu'}'^{\rm syn}$ is the spectral energy density of seed synchrotron photons, and $f_{\rm KN}$ is the KN scattering kernel, 
\begin{equation}\label{eq_fKN_this_work}
    f_{\rm KN} \left( \widetilde{x} \right)  =  \frac{3 }{8 \, \widetilde{x} } \int_{-1}^{+1} 
 d\mu (1-\mu)   \, \frac{\langle E_{\rm IC}'\rangle}{\gamma m_e c^2} \frac{\sigma_{\rm KN}[\widetilde{x}(1-\mu)] }{\sigma_{\rm T}} \, .
\end{equation}
with $\widetilde{x} =  v'/\widetilde{\nu}'$. Here $\widetilde{\nu}'$ is defined as the photon frequency beyond which KN effects become important when the seed photon is scattered by an electron with LF $\gamma$,
\begin{equation}\label{eq_nu_tilde}
    \widetilde{\nu}' (\gamma) = \frac{m_e c^2}{h} \frac{1}{\gamma} \, ,
\end{equation}
$\langle E_{\rm IC}'\rangle$ is the mean energy of the scattered photon, $\sigma_{\rm KN}$ is the KN cross-section, and $h$ is the Planck constant. In this work, we use the expression given by Eq. \ref{eq_Eic_average} for the mean scattered energy, which we use to calculate $f_{\rm KN}$ exactly (solid black curve in Fig.\,\ref{fig:fKN}). When using the approximation $\langle E_{\rm IC}'\rangle \approx \langle E_{\rm IC}' \rangle_{\rm T}/(1+\widetilde x)$ \citep[e.g.][]{Nakar+09} instead, it introduces small differences in the value of $f_{\rm KN}$ (dashed black curve in Fig.\,\ref{fig:fKN}).

The KN cross-section is suppressed for $\widetilde{x}>\widetilde{x}_{\rm KN}$ as $\sigma_{\rm KN}\propto E^{-1}$, whereas at $\widetilde{x}\ll \widetilde{x}_{\rm KN}$ it approaches $\sigma_{\rm T}$. A simple approximation is to consider a step function, with $\sigma_{\rm KN} = \sigma_{\rm T}$ for  $\widetilde{x}< \widetilde{x}_{\rm KN}$ and 
$\sigma_{\rm KN} = 0$  for $ \widetilde{x}'> \widetilde{x}_{\rm KN}$, which yields $f_{\rm KN}  = (1+ \widetilde{x})^{-1}$ for $\widetilde{x} < \widetilde{x}_{\rm KN} $ and $f_{\rm KN}  = 0$ otherwise. This most crude and commonly used approximation in many analytical works \citep[e.g.,][]{Nakar+09, Jacovich1+21, McCarthy-Laskar-24} is 
\begin{equation}\label{eq_fKN_step_approx}
     f_{\rm KN} \left( \widetilde{x} \right) = 
     \begin{cases}
       1 \, , & \widetilde{x} < \widetilde{x}_{\rm KN} \, ,\\
       0 \, , & \text{otherwise}  \,,
    \end{cases}
\end{equation}
where typically $\widetilde{x}_{\rm KN}=1$ and in some works $\widetilde{x}_{\rm KN}=0.2$ \citep{Yamasaki-Piran-22,Pellouin+24}. Fig. \ref{fig:fKN} shows that the step function is a very inaccurate approximation of $f_{\rm KN}$ as it severely underestimates the KN suppression for $\widetilde{x}<\widetilde{x}_{\rm KN}$. We also compare another approximation given in Eq.\,14 of \citet{Nakar+09}, which is much closer to the accurate result.

\begin{figure}
\centering
\includegraphics[width=1.\linewidth]{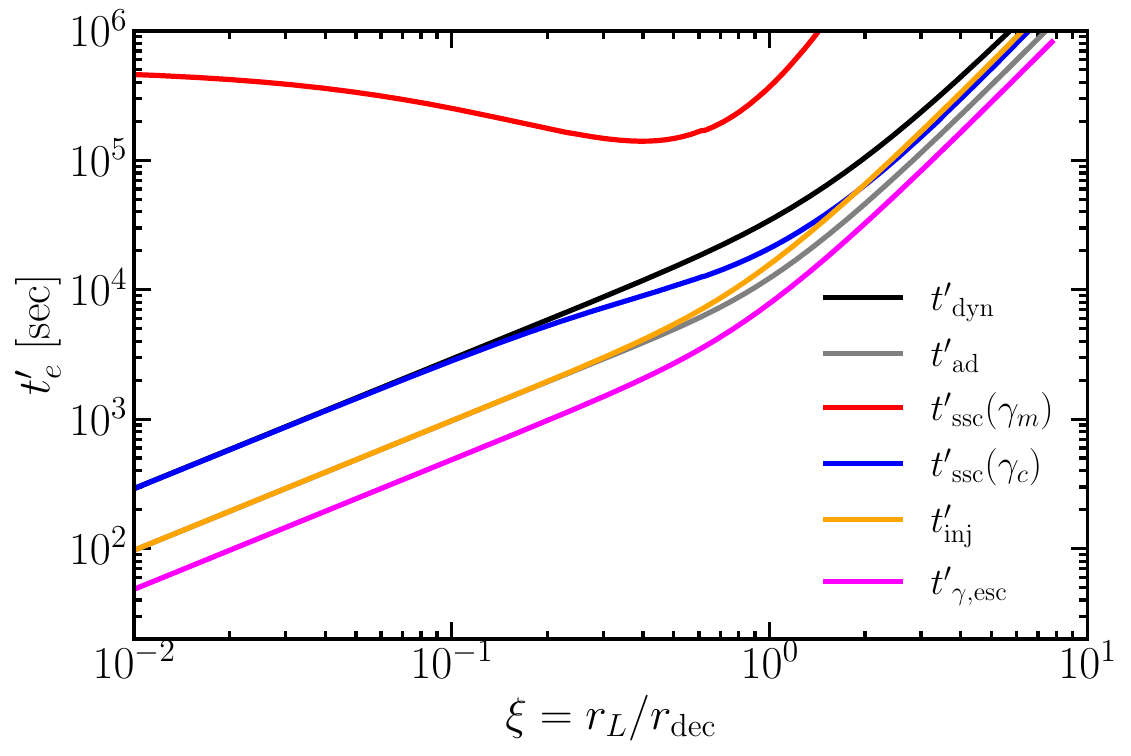} 
\vspace{-16pt} 
\caption{
Electron timescales calculated for the fiducial parameters 
$E_{k, \rm iso} =4 \times 10^{53} \, \rm erg$, $\Gamma_0 = 200$, $n_0 = 0.3 \, \rm cm^{-3}$, $k=0$, $\epsilon_e=10^{-1}$, $\epsilon_B=10^{-4}$, $p=2.3$,  $z=0.43$ ($d_L = 2.45 \, \rm Gpc$), $\gamma_{M} =  1 \times 10^8$. Where $r_L$ is the radius of the shell from which the emission arrives along the LOS at the apparent time $T$ and $r_{\rm dec}$ is the deceleration radius.
}
\label{fig:Comparison_timescale}
\end{figure}

%
%

\subsection{The Synchrotron-Self Compton Emission}

\subsubsection{Temporal evolution}
The evolution of the comoving photon distribution is described by the following continuity equation, 
that accounts for photon production from SSC emission, photon escape, and dilution of photon density 
due to expansion,
\begin{equation}
   \frac{\partial n'_{\nu'}(\nu, t')}{\partial t'} = \frac{ P_{\nu'}'^{\rm ssc}(\nu',t')}{h \nu'} 
   - \frac{n'_{\nu'}(\nu, t')}{t'_{\rm esc}} - \frac{n'_{\nu'}(\nu, t')}{t'_{\rm ad}} \, ,
\end{equation}
where $n'_{\nu'}(\nu') = \partial n'/\partial\nu'$ is the photon number density per unit frequency, $ P_{\nu'}'^{\rm ssc} \equiv P_{\nu'}'^{\rm syn} + P_{\nu'}'^{\rm IC}$ is the SSC radiated power per unit volume per unit frequency, and $t_{\rm esc}' = 2\Delta'/c$ is the photon escape timescale \citep{Fukushima+17}, where the factor of 2 accounts for escape from the two surfaces of the shell when approximated as a slab of comoving radial width $\Delta'$. In general, the above equation is non-linear since the IC spectral radiated power depends on the photon distribution at any given time $t'$. However, when considering this equation over timescales smaller than the shortest timescale in the problem, e.g. the photon escape time, as demonstrated in Fig.\,\ref{fig:Comparison_timescale} for a set of fiducial afterglow model parameters, then the SSC spectral radiated power can be approximated to be constant over $\Delta t' \ll \min( t'_{\rm esc}, t'_{e,\rm cool}, t'_{\rm ad} )$. Therefore, we discretize the evolution of the radiation field in the emission region over steps in radius $\Delta r = \Gamma(r)\beta(r)c\Delta t'(r)$ where properties of the emission are considered to be constant between $r$ and $r+\Delta r$. In this case, the spectral energy density, $u'_{\nu'}=h\nu'\,n_{\nu'}'$, of the total radiation field over time $\Delta t'$ can be expressed as (see appendix \ref{Appendix:photon_cont_eq} for more details)
\begin{multline}\label{eq_uSSC_sol}
    u_{\nu'}'(r) \approx 
    %
    \left[P_{\nu'}'^{\rm ssc}(r)\Delta t' + u'_{\nu'}(r-\Delta r) \right] \, \times \\ 
    \exp \left[ - \Delta t'(r) \left( t'_{\rm esc}(r)^{-1} + t'_{\rm ad}(r)^{-1} \right ) \right] \, ,
\end{multline}
where the first term on the R.H.S is the contribution due to the newly produced radiation 
over time step $\Delta t'$ and the second term is the radiation field in the emission region at 
radius $r-\Delta r$, with spectral energy density $u'_{\nu'}(r-\Delta r)$, both reduced by photon 
escape and diluted by adiabatic expansion. Since the photon escape time varies with radius, so 
does the interval $\Delta t'$ which is taken to be a constant fraction of $t_{\rm esc}'(r)$. 

\subsubsection{Radiated Power}
The synchrotron radiated power per unit volume per unit frequency produced by a distribution of electrons can be calculated by integrating the emission from a single electron over the entire distribution, such that
\begin{equation}\label{eq_jSYN}
        P_{\nu'}'^{\rm syn} (\nu')
        = \int_{ 1 }^{\gamma_{M}} 
        d\gamma \, n_e'(\gamma)
        \, P_{\nu', e}'^{\rm syn} (\nu', \gamma) \,  ,
\end{equation}
where the synchrotron power per unit frequency emitted by a single electron accounting for the contribution of all pitch angles is given by \citep[e.g.][]{Rybicki-Lightman-79}
\begin{equation}\label{eq_SYN_power_nuc}
P_{\nu', e}'^{\rm syn}(\nu') = \frac{\sqrt{3} e^3 B'}{m_e c^2} \,  K_{\rm syn} \left(\frac{\nu'}{\nu_c'}\right), 
\quad 
\text{with}  
    \quad  
\nu_c' = \frac{3}{2} \frac{e B'}{2\pi m_e c} {{\gamma}^2} \, ,
\end{equation}
where $K_{\rm syn}$ is the synchrotron kernel for a single electron given by $K_{\rm syn}(x_c) =  \, (x_c/2) \int_{0}^\pi d\alpha \sin\alpha \int_{(x_c/\sin \alpha)}^{\infty} dt K_{5/3} (t)$. 
The radiated power per unit volume per unit frequency of IC scattered synchrotron photons by the same electron distribution is given by 
\begin{equation}\label{eq_jIC}
    P_{\nu'}'^{\rm IC}
    =
    \frac{3}{4} c \, \sigma_{\rm T} \nu_{\rm IC}'
    \int_{\gamma_{\rm 0,IC }}^{\gamma_{M}} d\gamma \frac{n_e'(\gamma)}{\gamma^2}
    \int_{\nu_{\rm min}'}^{\nu_{\rm max}'} d\nu' \frac{ u_{\nu'}'^{\rm syn} }{\nu'^2} F_c(q,\Gamma_e) \, ,     
\end{equation}
where $\gamma_{\rm 0,IC } = \max\left(1, \frac{h \nu_{\rm IC}' }{m_e c^2}\right)$ and the above expression is valid only for an isotropic and homogeneous photon distribution. The last term is the IC's kernel which includes Klein-Nishina effects $F_{\rm C}(q,\Gamma_e)=2q \log(q) + (1+2q)(1-q) + \frac{1}{2} \frac{(\Gamma_e q)^2}{1 + \Gamma_e q} (1- q) \, ,$
with $\Gamma_e = \frac{ 4 h }{m_e c^2} \nu' \gamma$  and $q=\frac{ h \nu_{\rm IC}'}{\Gamma_e ( m_e c^2  \gamma - h \nu_{\rm IC}' )}$ \citep{Blumenthal-Gould-70}. From relativistic kinematics $q$ is constrained to be in the range of $1/(4\gamma^2) \leq q \leq 1$, and therefore using this condition, the photon integration limits are
\begin{equation}\label{eq:IClimits_epsilon}
 \nu_{\rm min}' = \frac{1}{4\gamma^2} \nu_{\rm max}'
 \qquad \text{and} \qquad
 \nu_{\rm max}' = \frac{\nu_{\rm IC}'}{1- \frac{h \nu_{\rm IC}'}{ \gamma m_e c^2} } \, .
\end{equation}

\begin{figure}
\centering
\includegraphics[width=1.\linewidth]{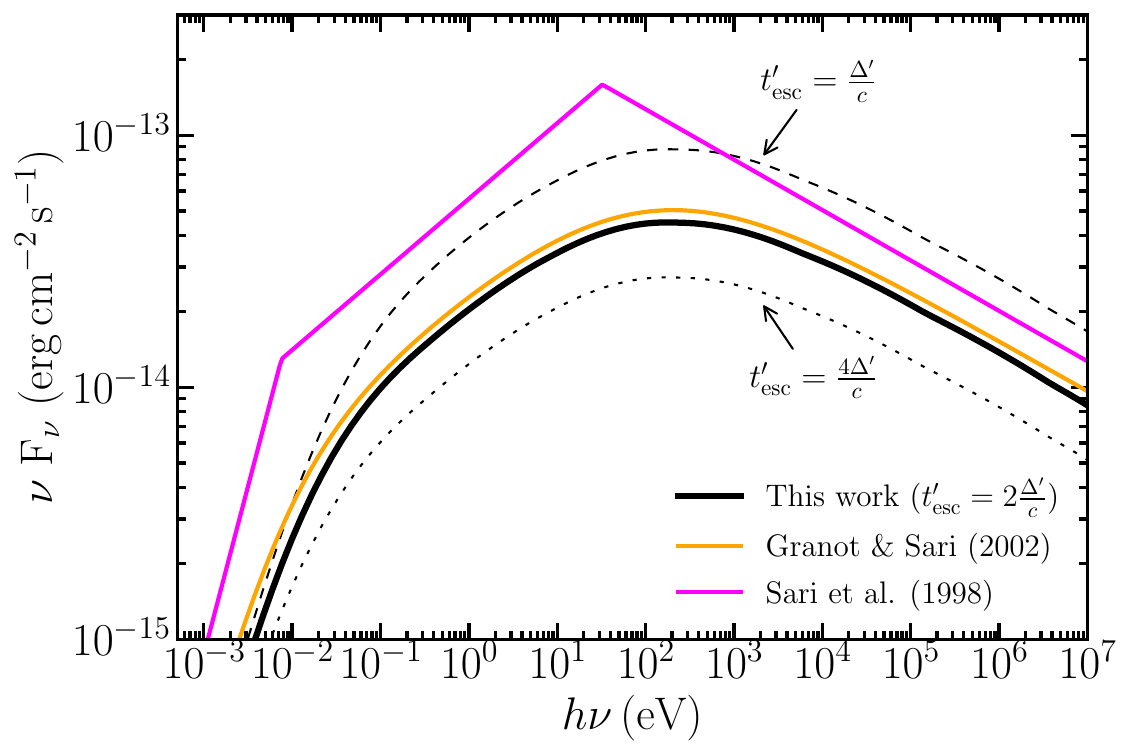} 
\vspace{-16pt} 
\caption{Comparison of slow-cooling synchrotron spectrum from different models with that obtained in this work in the case when $Y_{\rm ssc}(\gamma) \ll 1$, for the following model parameters:
$E_{\rm k, iso} = 10^{52} \, \rm erg$, 
$\Gamma_0 = 200$, 
$n_0 = 10^{-2} \, \rm cm^{-3}$, 
$k=0$, 
$\epsilon_e = 10^{-1}$, 
$\epsilon_B = 8\times 10^{-2}$, 
$p=2.4$, 
$T = 10^5 \, \rm s $, 
$z=1 \, (6.8 \, \rm Gpc)$.
The black curve shows the spectrum from our model after EATS integration, which is compared with the 3D shocked fluid volume and EATS integrated spectrum from \protect\cite{Granot-Sari-02} (orange) and the LOS spectrum from \protect\cite{Sari+98} (magenta). Dashed and dotted lines are calculated using our model but assuming shorter and longer escape timescale, respectively. This verifies that our chosen escape time (solid black) is optimal.
}
\label{fig:EATS_LOS_COMPARISON}
\end{figure}

\subsection{The Observed Spectrum}\label{sec::Observed_spectra}

Due to relativistic and light travel time effects, photons emitted from a spherical shell at different radii but from the same 
angular distance $\theta$ away from the LOS, or from the same radius but from different $\theta$, arrive at 
the observer at different times $T$. The locus of points over different radii and $\theta$ forms an 
\textit{equal arrival time surface} (EATS), as given by
\begin{equation}\label{eq_EATS_condition}
    T_{z} \equiv \frac{T}{1+z} = t(r) - \frac{r\mu}{ c} \, ,
\end{equation}
from which emission arrives at the observer at a given time $T$, where $T_z$ is the arrival time of photons 
in the cosmological rest-frame of the central engine, $\mu = \cos \theta$, and $t$ is the lab-frame time, 
as given by
\begin{equation}\label{eq_tlab}
    t(r) = \int_0^r \frac{d r' }{\beta(r') c} \,,
\end{equation}
for the radiating shell to arrive at radius $r$. 
The observed spectrum is composed of contributions from Doppler boosted radiation emitted at different 
radii and from different parts of the shell (mainly that originating inside the $1/\Gamma(r)$ beaming 
cone around the LOS), which can be calculated from \citep[e.g.][]{Gill-Granot-18b}
\begin{equation} \label{eq_EATS_flux}
    F_{\nu}(\nu, T) = 
    \frac{(1+z)}{8\pi d_L^2}
    \int_{\mu_{\rm j}}^1 d\mu \,
    \delta_D^3(\mu,r) \, L_{\nu'}'(\nu', r)   \, ,       
\end{equation}
where $\mu_j= \cos \theta_j$ and $\theta_j$ is the half-open angle of the jet, 
$\delta_D = (1+z)\nu/\nu' = \big[\Gamma(r)\{1-\beta(r) \mu \}\big]^{-1}$ is the Doppler factor. 
The comoving spectral luminosity, $L_{\nu'}'$ that escapes from the emitting region is given by 
\begin{equation}\label{eq_Lnu_escape}
     L_{\nu'}'(\nu', r) = 4\pi \, r^2 \, \Delta'(r) \, 
     f_{\rm esc} (r)
     \frac{ u_{\nu'}'(\nu',r) }{\Delta t' (r)}
     \,,
\end{equation}
where (see appendix \ref{Appendix:photon_cont_eq} for more details)
\begin{equation}\label{eq_fesc}
    f_{\rm esc}(r) =  \frac{1 - \exp(-\Delta t'/t_{\rm esc}') }{\exp(-\Delta t'/t_{\rm esc}')}
\end{equation}
is the fraction of photons that escape from the emitting region over time $\Delta t'$.



\begin{figure}
\centering
\includegraphics[width=1\linewidth]{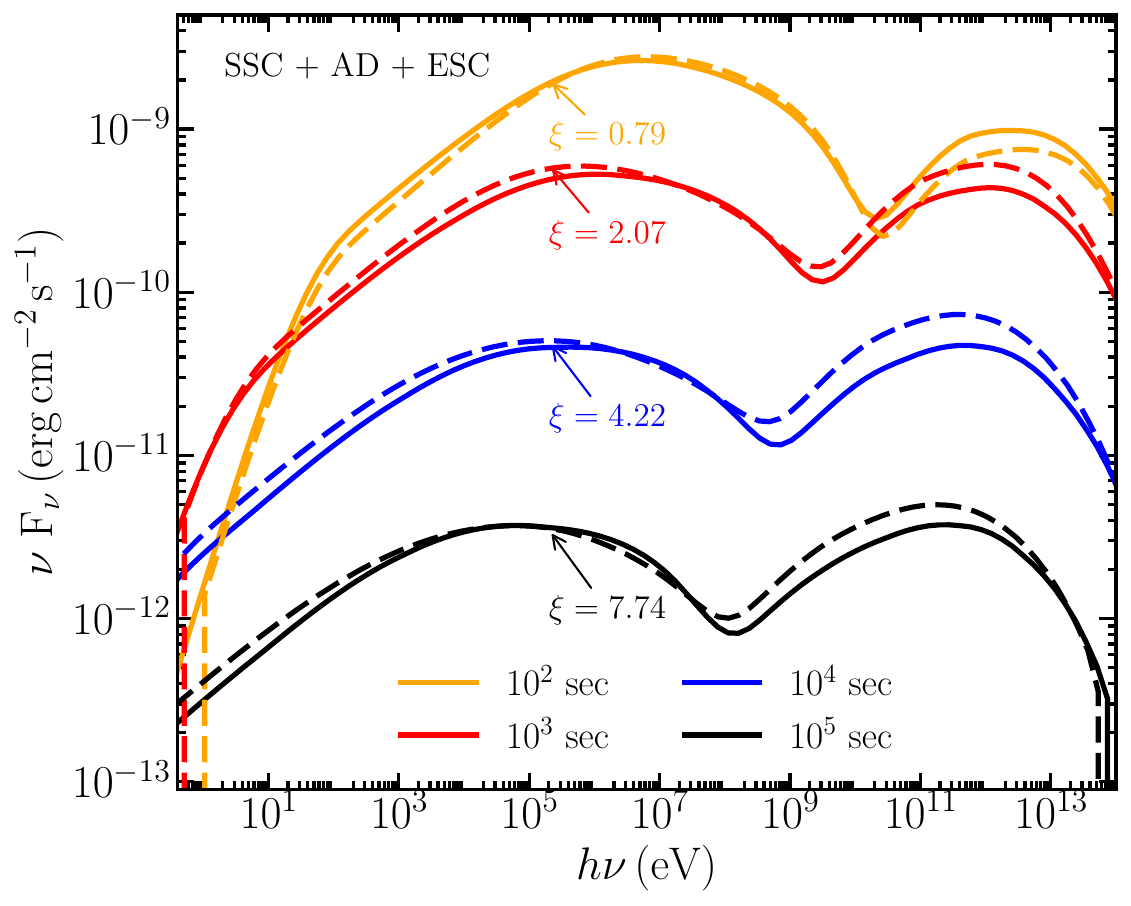} 
\vspace{-16pt} 
\caption{
Comparison between our model (solid lines) and results obtained using a numerical kinetic code 
(dashed lines). The parameters used to calculate the spectra are
$E_{k, \rm iso} =4 \times 10^{53} \, \rm erg$, $\Gamma_0 = 200$, 
$n_0 = 0.3 \, \rm cm^{-3}$, $k=0$, 
$\epsilon_e=10^{-1}$, $\epsilon_B=10^{-4}$, 
$p=2.3$, $\gamma_{M} =  1 \times 10^8$, 
$z=0.43$ ($d_L = 2.45 \, \rm Gpc$). The different spectra are shown at different observer-frame times $T$ that corresponds to different $\xi=r_L/r_{\rm dec}$, where $r_L$ is the radius from which emission arrives at the observer at time $T$ along the LOS. Both models include SSC, adiabatic cooling (for particles) and density dilution (particles and photons), and escape of radiation from the emission region.
}
\label{fig:Comparison}
\end{figure}

\section{Model Validation \& Improvements}\label{sec::MODEL VALIDATION}

\begin{figure*}
\centering
\includegraphics[width=0.48\textwidth]{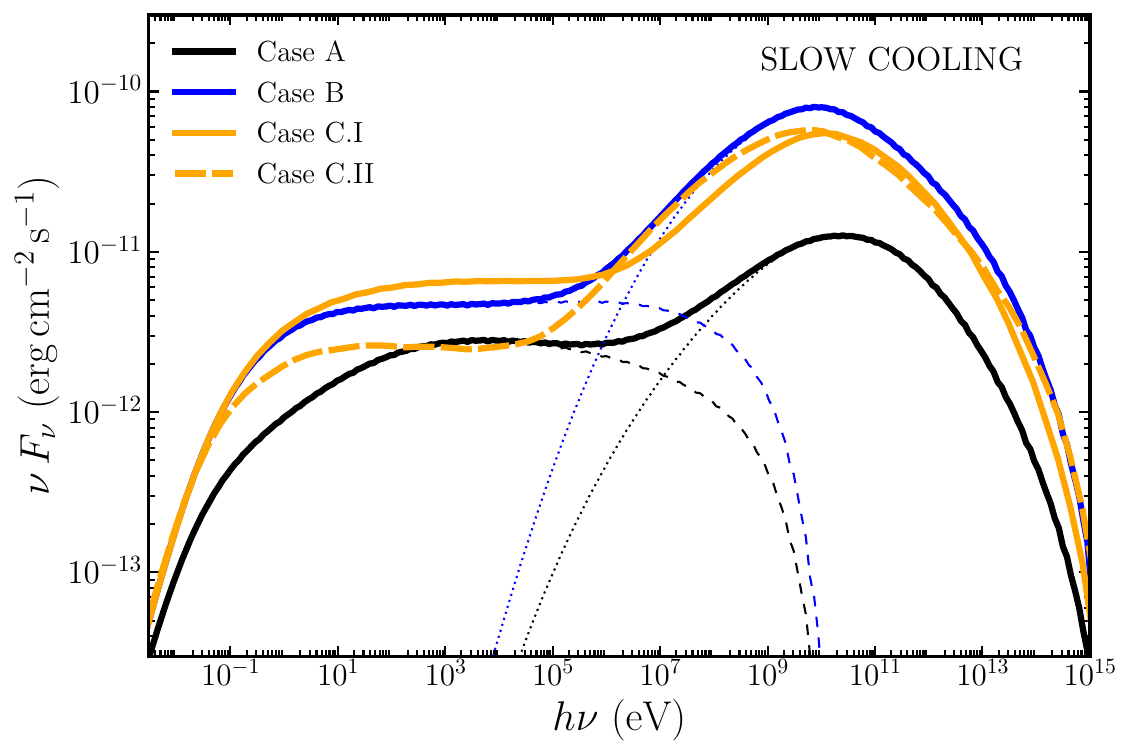}
\includegraphics[width=0.48\textwidth]{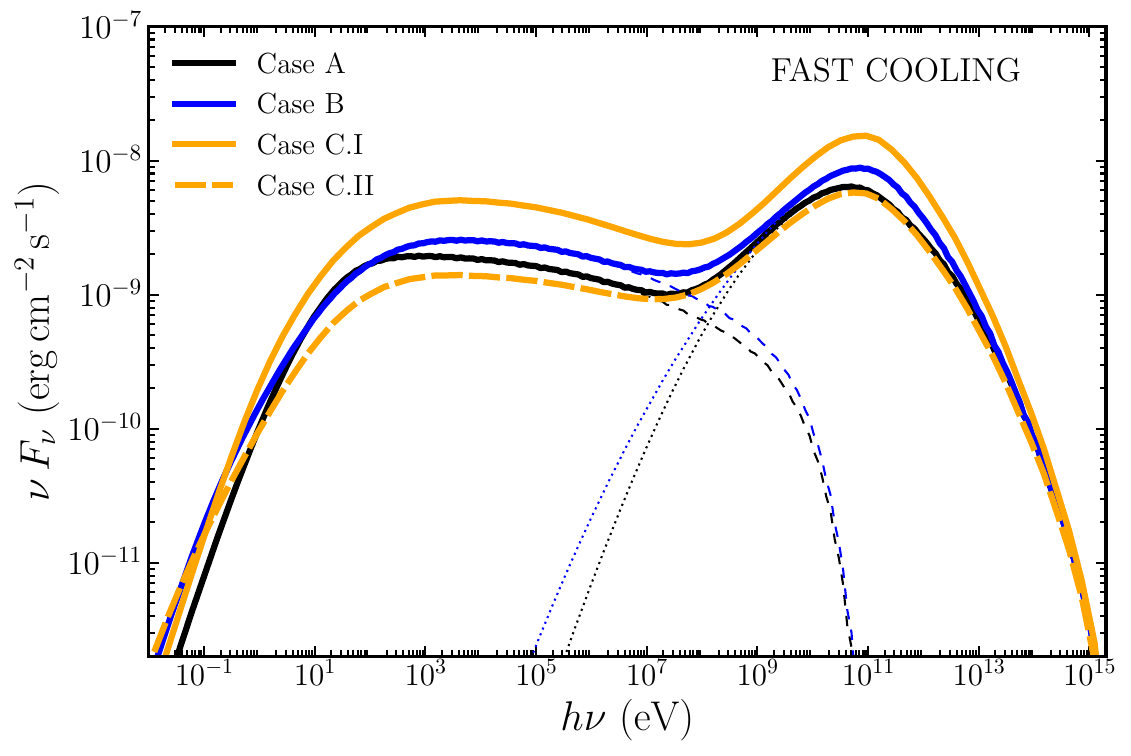}
\caption{
Comparison of the resulting spectra obtained from the approach used in this work, considering different effects A) SSC, photon escape and adiabatic expansion (black lines), B) SSC and photon escape (blue lines), and C) SSC with shorter (solid yellow lines) and larger (dashed yellow lines) photon escape timescales. Thin dashed lines represent the synchrotron component and dotted lines represent the IC component. We use the following set of parameters $\Gamma_0 = 1 \times 10^{3}$, $n_0 = 1 \, \rm cm^{-3}$, $k=0$, 
$\epsilon_e = 10^{-1}$, $p=2.4$ and $z=1$. For slow cooling we set 
$\epsilon_B = 10^{-4}$ and $E_{k, \rm iso} = 1.08 \times 10^{54} \, \rm erg$ at $T = 2.07 \times 10^4\, \rm sec $. For fast cooling we set 
$\epsilon_B = 10^{-3}$ and $E_{k, \rm iso} = 4.04 \times 10^{54} \, \rm erg$ at $T = 5.97 \times 10^{2}\, \rm sec $. 
}
\label{fig_spectra_effects}
\end{figure*}

\subsection{Comparison with Synchrotron Only Analytic Models}

We first validate the accuracy of the synchrotron spectral component in our model by comparing it with the standard analytic model introduced by \citet[][hereafter, SPN98]{Sari+98} and the improved and widely used prescription developed by \citet[][hereafter, GS02]{Granot-Sari-02}. The SPN98 model is an analytic framework in which afterglow synchrotron spectrum arises from shock-accelerated electrons injected with a power-law energy distribution by the ultra-relativistic blast wave. The emission is calculated only along the LOS without integration over the EATS and features different power-law segments joined sharply at characteristic frequencies. In comparison, the GS02 model presents a more realistic approach that incorporates particle adiabatic cooling downstream of the shock and performs EATS integration over the entire 3D shocked volume behind the shock. It also provides a prescription for obtaining smoother spectral breaks between the different power-law segments.

Fig.\,\ref{fig:EATS_LOS_COMPARISON}
compares the slow-cooling synchrotron spectrum, with $Y_{\rm ssc} \ll 1$, from these two models and that obtained from our model. It shows excellent agreement between our results and the spectrum obtained from the GS02 model, and confirms that the thin shell approximation used in this work is sufficient to explain the synchrotron component. Both of these models do not consider SSC emission and in particular effects on the seed synchrotron spectrum due to KN effects. We make comparisons with numerical models next that do include such effects.

\subsection{Comparison with Numerical Models}

Here we consider numerical models that produce the VHE TeV emission with SSC and include KN effects on both the seed and IC scattered emission. In particular, we use the numerical model from Gill et al. (2025, in preparation) that performs kinetic simulations of the interactions between photons and electrons, including adiabatic cooling of electrons, density dilution of both species due to expansion, and escape of radiation from the emission region. The observed emission is obtained by performing an integration over the EATS.

Fig.\,\ref{fig:Comparison} presents a comparison between the broadband numerical spectrum and that obtained 
from our model, for different observer-frame times $T$ corresponding to different shell radii $\xi=r_L/r_{\rm dec}$, 
where $r_L$ is the radius of the shell from which the emission arrives along the LOS at the apparent time $T$.
The results from our semi-analytical model are in good agreement with the numerical model over the evolution of the afterglow phase, both before ($\xi<1$) and after ($\xi>1$) the deceleration radius. 
Specifically, in the synchrotron component, both approaches exhibit very similar spectral shape and flux levels. 
Only minor differences appear near and below the location of the peak frequency. 
When comparing the SSC component at high energies, our results agree well in spectral shape, however, small 
differences arises in the flux level between the two approaches. The flux in our model is about a factor of 
$\approx 1.3$ higher than that predicted by the kinetic approach before the deceleration radius, whereas after 
the deceleration radius our model predicts about a factor of $\approx 1.25-1.4$ smaller flux.
The good agreement between the two results confirms the validity of our semi-analytic treatment in describing the time-dependent evolution of the broadband 
afterglow emission.

\begin{table*}
\centering
\setlength{\tabcolsep}{6pt} 
\renewcommand{\arraystretch}{1.2} 
\begin{tabular}{llccc|ccc}
\toprule
\multicolumn{2}{c}{} & \multicolumn{3}{c}{\textbf{SLOW COOLING}} & \multicolumn{3}{c}{\textbf{FAST COOLING}} \\
\cmidrule(lr){3-5} \cmidrule(lr){6-8}
\multicolumn{1}{c}{\textbf{MODEL}} & \multicolumn{1}{c}{\textbf{CASE}} & \( \gamma_c \, [10^3] \) & \( \gamma_0 \, [10^{11}]\) & \( Y_{
                   \rm Th } \)
                    & \( \gamma_c \, [10^2]\) & \( \gamma_0 \, [10^9]\) & \( Y_{
                   \rm Th } \) \\
\midrule

\multirow{3}{*}{\textbf{This work}} 
  & \textbf{A}: SSC + ESC + AD  
        & $5.25 $ & $0.02$ & $9.7$
        & $1.48$ & $1.60$ & $20.5$ \\
  & \textbf{B}: SSC + ESC       
        & $2.49 $ & $0.24$ & $20.2$ 
        & $1.65$ & $1.09$ & $17.1$ \\
  
\midrule

\multirow{1}{*}{\textbf{NAS09}} 
  & SSC             & $3.36$ & $2.12$ & $15.2$
                    & $1.00$& $25.9$ &  $31.1$ \\

\bottomrule
\end{tabular}
\caption{Characteristic values calculated using different physical effects, obtained from our model at LOS radius and from the NAS09 formalism. We also include predictions from an analytic which includes radiation escape and adiabatic expansion. 
\newline
\textbf{Notes:} \( \gamma_c = \gamma_c^{\rm syn} / [1 + Y_{\rm ssc}(\gamma_c)] \) – cooling Lorentz factor; 
\( \gamma_0 \) satisfy \(Y_{\rm ssc}(\gamma_0)=1 \); 
\( \eta_{\text{rad}} = \min\big[1, \left( \frac{\gamma_m}{\gamma_c}  \right) ^{p-2} \big] \) – SSC radiative efficiency; \( Y_{\rm Th} = \sqrt{\eta_{\rm rad}\frac{\epsilon_e}{\epsilon_B}} \) – Thomson Compton-Y parameter.
}

\label{tab:Comparative_NAS09}
\end{table*}

\subsection{Importance of Adiabatic Cooling and Photon Escape}

All analytical works on this topic generally do not include the effect that adiabatic cooling has on the particle distribution, and consequently on the SSC spectrum. In addition, all such works also make the assumption of an infinitely thin shell, i.e. whatever radiation field is produced at any given time in the fluid frame is radiated instantaneously. Neglect of both effects lead to significant changes in the model spectrum, as we demonstrate here, and as a result, incorrect inference of the physical model parameters when comparisons are made with observations. In the following, we consider three different cases:

\begin{enumerate}
    \item \textbf{Case A}: Both photon escape and adiabatic cooling and dilution are considered. 
    \item \textbf{Case B}: Only photon escape and no adiabatic cooling and dilution.
    Adiabatic cooling is neglected by imposing $t_{\rm ad}' \gg \Delta t'$, whereby we set $t_{\rm ad}'$ to some arbitrarily large time.
    \item \textbf{Case C}: No adiabatic cooling and dilution, but photon escape at two different rates: (C.I) Escape of photons is made rapid by artificially shortening the escape time by half; (C.II) photon escape rate is made slower by making the escape time longer by a factor of 2.
\end{enumerate}

In Fig.\,\ref{fig_spectra_effects}, we calculate the SSC spectra for all these cases in the slow (left panel) and fast (right panel) cooling regimes. We discuss these cases in more detail below.

\subsubsection{No Adiabatic Cooling and Dilution}
Adiabatic expansion of the emitting region acts to dilute both the photon and electron number densities, and it also cools the electrons. The obvious outcome is a larger overall normalization of the spectrum, but more importantly case B shows a markedly larger $Y_{\rm ssc}$ in the slow-cooling regime. Since the synchrotron spectral radiated power scales with electron density, $P_{\nu',\rm syn}'\propto n_e'(\gamma)$, and the IC spectral radiated power scales even strongly, $P_{\nu',\rm IC}'\propto n_e'(\gamma)^2$ 
it leads to a larger $Y_{\rm ssc}$. The enhanced cooling now reduces the cooling LF of electrons, $\gamma_c \propto [1 + Y_{\rm ssc}(\gamma_c)]^{-1}$, and moves the cooling break to lower energies. A slightly harder spectrum above $\nu_c$ is also obtained.

The spectrum is not so significantly affected in the fast-cooling case, apart from having a slightly larger overall normalization. Only small differences with respect to case A arise due to SSC cooling dominating completely over adiabatic cooling.

\subsubsection{Photon Escape}
In many analytic works no light travel time effects for the radiation field are included, and all photons from the entire emitting region arrive at the observer at the same time. In fact, it takes a finite amount of time, i.e. $\sim\Delta/c$ in the lab-frame, for the radiation field to traverse the width of the emission region.
Similarly to adiabatic expansion, but more importantly, photon escape acts as a dilution effect on the photon energy density in the emitting region (see Eq. \ref{eq_uSSC_sol}). 
A shorter photon timescale reduces the fraction of photons that remain within the emission region over a dynamical timescale, which reduces the Compton-Y parameter and increases the observed flux normalization. To assess how the rate of radiation escape changes the spectra, we compare cases B and C in Fig.\,\ref{fig_spectra_effects}.
    
    %
    %
When comparing the three cases in the slow-cooling regime, the modification to the SSC peak flux and the IC spectrum are rather minimal. This also means that the photon escape timescale produces a subdominant effect on the IC spectrum when compared to that produced by adiabatic cooling and dilution. The normalization of the synchrotron spectrum is changed, however, which also affects the Compton-Y parameter. A faster rate of photon escape (C.I) naturally yields a larger normalization of the synchrotron component over a 4 times slower rate of escape (C.II). The slightly larger value of $Y_{\rm ssc}$ in C.II also shifts the cooling break to lower energies, which is reflected in the shift of the spectral peaks to slightly lower energies.

While large spectral changes are seen in the slow-cooling case, the overall spectrum in the fast cooling regime is not modified so drastically. Since particles radiate away their energy on much shorter timescales, adiabatic cooling is not the dominant effect anymore. Faster radiation escape simply leads to a larger overall normalization of the spectrum. 
    
The example spectra shown in Fig.\,\ref{fig_spectra_effects} clearly demonstrate the effect of accurately accounting for adiabatic cooling \& dilution as well as rate of photon escape. We find that neglecting either of the two effects leads to significant deviation of the SSC spectra, particularly for the slow cooling regime. 

\subsection{Comparison with \protect\cite{Nakar+09}}

Here we compare our model with the widely used analytic formalism of \citet[][ hereafter, NAS09]{Nakar+09}, which is the foundation for most current analytic models of SSC emission with KN corrections. The goal is to show that due to the simplifications in the NAS09 formalism its predicted spectrum deviates in important ways from that obtained from our semi-analytic approach. 

\begin{figure}
    \centering
    \includegraphics[width=0.48\textwidth]{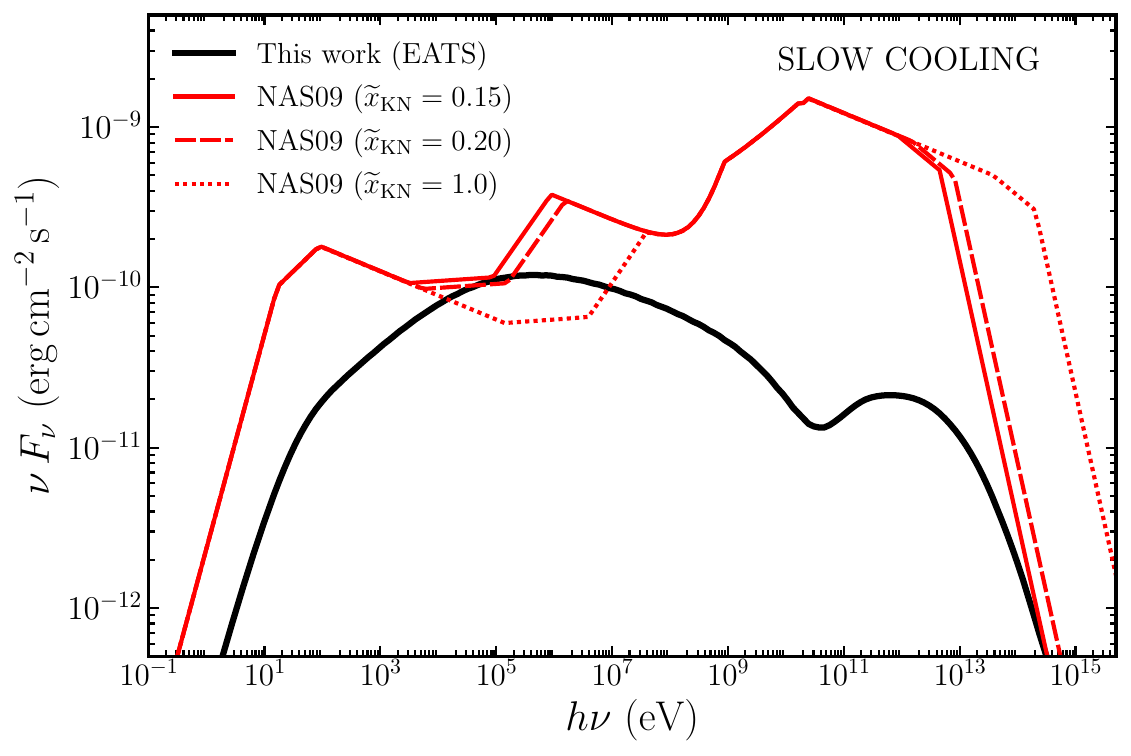}
    \caption{LOS analytic spectrum from NAS09 for the different values of $\widetilde{x}_{\rm KN}$ and for the following set of model parameters: $E_{k, \rm iso} = 1 \times 10^{53}\, \rm erg$, $\Gamma_0 = 200 $, $n_0 = 0.1 \, \rm cm^{-3}$, $k=0$, $\epsilon_e = 10^{-1}$, $\epsilon_B = 10^{-3}$, $p=2.3$ and $z=1$ at $T = 200 \, \rm s $. EATS integrated spectrum from our model (solid black) is shown for comparison.
   }
    \label{fig_SSC_xKN_data}
\end{figure}

Fig. \ref{fig_SSC_xKN_data} shows a comparison of the spectra obtained using the NAS09 formalism along the LOS with different values of $\widetilde{x}_{\rm KN}$. Many works that make use of the NAS09 formalism set $\widetilde{x}_{\rm KN} = 1$ or $\widetilde{x}_{\rm KN} = 0.2$, where the latter arguably yields results closer to the more accurate solution \citep{Yamasaki-Piran-22}. Alternatively, one can use $\widetilde{x}_{\rm KN} = 0.15$ that approximates $f_{\rm KN}$ at its midpoint. As expected, smaller values of $\widetilde{x}_{\rm KN}$ move the energy scale above which KN effects modify the spectrum to lower energies. While only small differences are observed between $\widetilde{x}_{\rm KN} = 0.15$ and $\widetilde{x}_{\rm KN} = 0.2$, the resulting spectra for $\widetilde{x}_{\rm KN} = 1$ exhibits a broader and deeper valley in the synchrotron component, as well as an IC component that extends to higher energies by at least one order of magnitude in frequency. 

Comparing with our approach (solid black), and specifically examining the ratio of peak fluxes, it is clear that the two models predict significantly different spectra. While NAS09 model predicts $Y_{\rm ssc} \sim \sqrt{(\gamma_c/\gamma_m)^{2-p} \, \epsilon_e/\epsilon_B} \sim 10 $ at the peak, our model yields $Y_{\rm ssc} < 1$. This discrepancy between the two approaches produces peak frequencies that differ by almost two orders of magnitude, which can result in obtaining very different sets of model parameters when comparing with observations.

\begin{figure*}
\begin{minipage}[b]{0.50\linewidth}
\centering
\includegraphics[width=1.\linewidth]{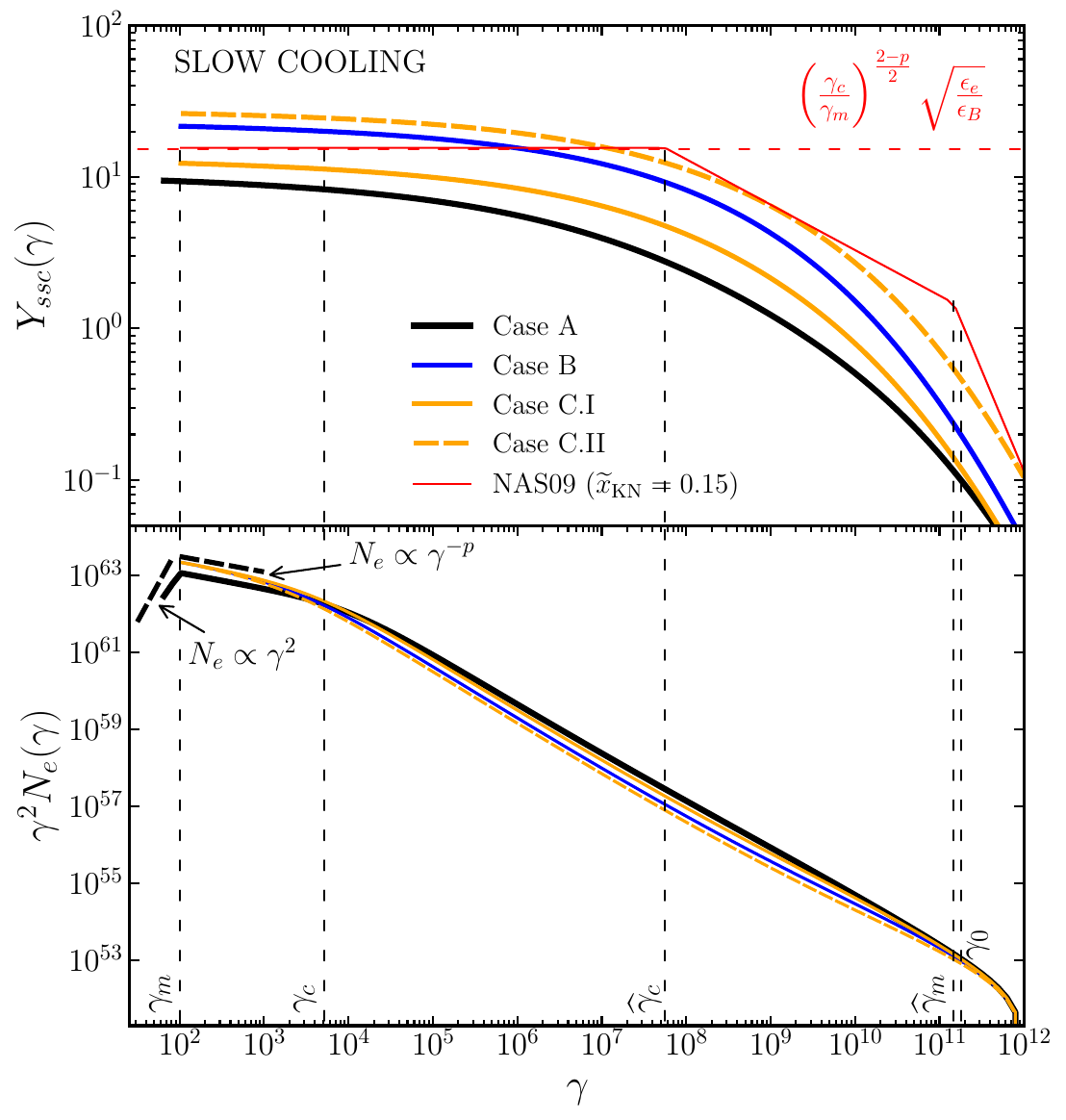} 
\end{minipage}\hfill 
\begin{minipage}[b]{0.50\linewidth}
\centering
\includegraphics[width=1.\linewidth]{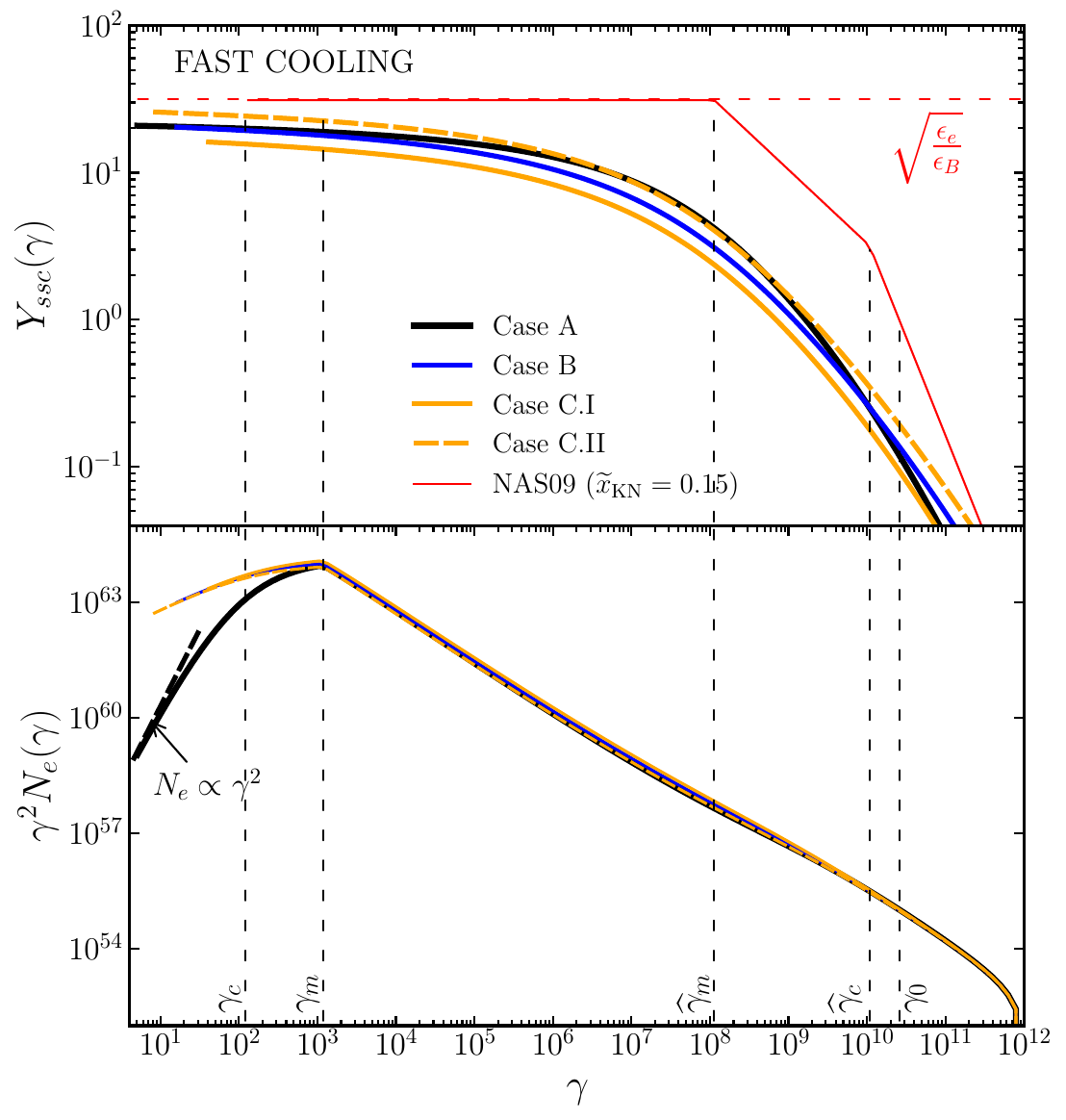} 
\end{minipage} 
\vspace{-16pt} 
\caption{
Comparison between our semi-analytic model and the analytic model of NAS09 with $\widetilde{x}_{\rm KN}=0.15$, in the slow and fast (case I in NAS09) cooling regimes. 
We use the following set of parameters $\Gamma_0 = 1 \times 10^{3}$, $n_0 = 1 \, \rm cm^{-3}$, $k=0$, 
$\epsilon_e = 10^{-1}$, $\epsilon_B = 10^{-3}$, $p=2.4$, $z=1$ ($d_L =6.8 \, \rm Gpc$), $\gamma_{M}=10^{12}$. 
The upper panels show the Compton-Y 
parameter and the lower panels show the 
electron distribution, both are calculated at $r_L$. Dashed lines indicate the expected power-law behavior of the electron distribution given in Eqs. \ref{eq_eD_fast} and \ref{eq_eD_slow}, as well as the characteristic electron LFs. 
\textbf{Left:} Slow cooling case, with $\gamma_m/\widehat{\gamma}_m = 2 \times 10^{-9}$ yielding $E_{k, \rm iso} = 2.48 \times 10^{59} \, \rm erg$ at $T = 2.01 \times 10^{8}\, \rm sec $. 
\textbf{Right:} Fast cooling case, with $\gamma_m/\widehat{\gamma}_m = 1 \times 10^{-5}$ yielding, $E_{k, \rm iso} = 5.15 \times 10^{61} \, \rm erg$ at $T = 4.08 \times 10^{6}\, \rm sec $.  
\textbf{Note:} Unrealistic physical parameters were chosen in this figure to achieve a large dynamical range between the different breaks to demonstrate the asymptotic behavior within them.
}
\label{fig_Y_ne_comparison}
\end{figure*}

\begin{figure*}
\centering
\begin{minipage}[b]{0.50\linewidth}
\centering
\includegraphics[width=1.\linewidth]{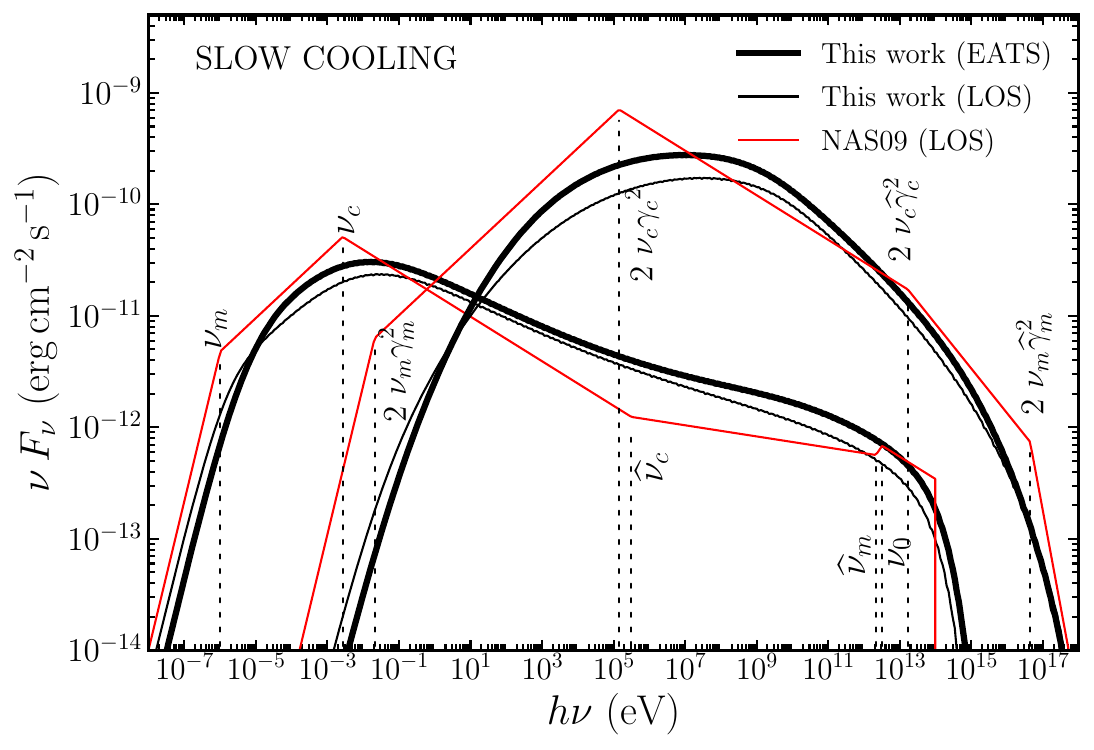} 
\end{minipage}\hfill 
\begin{minipage}[b]{0.50\linewidth}
\centering
\includegraphics[width=1.\linewidth]{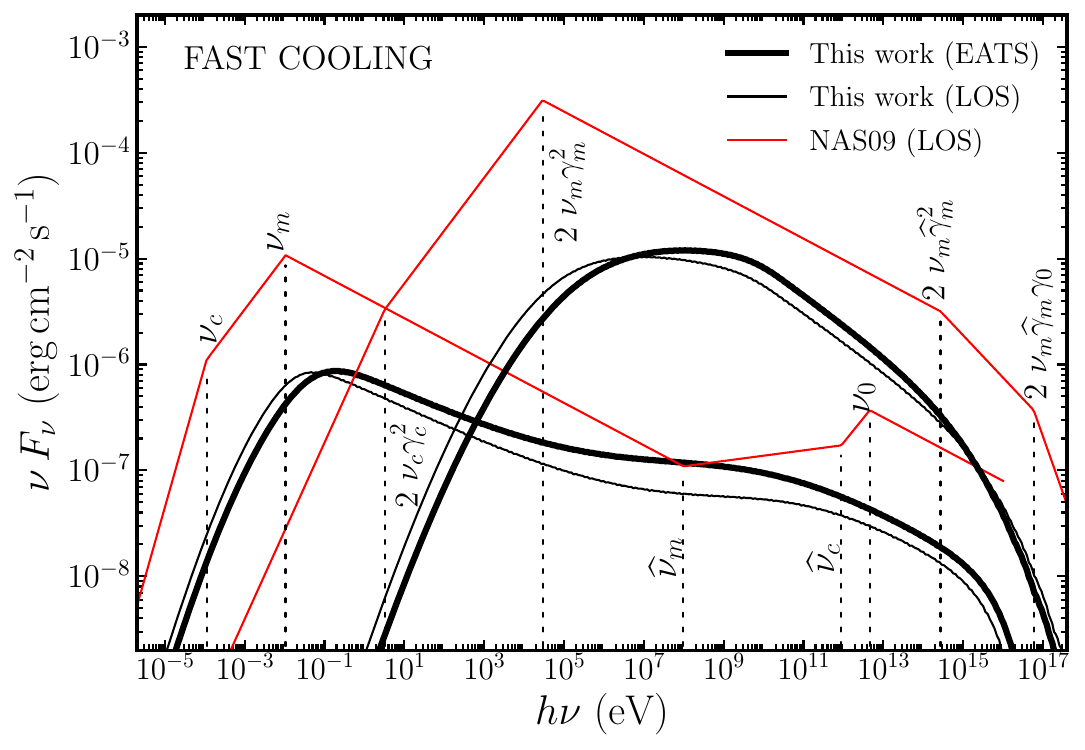} 
\end{minipage} 
\vspace{-16pt} 
\caption{
    Comparison between the spectra obtained from our model (black lines) and those the obtained from the \citet{Nakar+09} formalism (red lines), using the same parameter set as Fig. \ref{fig_Y_ne_comparison}. Thin lines represent solutions considering only the emission along the LOS and thick lines represent the spectra obtained by EATS integration. 
    \textbf{Note:} Unrealistic physical parameters were chosen in this figure to achieve a large dynamical range between the different breaks to demonstrate the asymptotic behavior within them.
}
\label{FIG_COMPARISON_NAKAR}
\end{figure*}

In Fig.\,\ref{fig_Y_ne_comparison}, we compare the Compton-Y parameter and particle distribution in the slow (left-panel) and fast (right panel) cooling regimes. In order to show the asymptotic behavior of our solution, and to make the comparison easier with the power-law solution of NAS09, some of the model parameters assume extreme values that are never realized in GRB afterglows. In general, we find that our solution of $Y_{\rm ssc}$ agrees with that of NAS09 and represents a smoother version of the power-law approximation obtained in NAS09. However, there are some differences in the normalizations that arise due to several simplifying assumptions in the analytic treatment of NAS09 and the additional effects included in our model. In particular, we find that our model yields lower values of the $Y_{\rm ssc}$. In the Thomson regime, these are within a factor of $2$ to that obtained from NAS09, but in the KN regime the difference are significant. The comparison of the different break energies and normalizations are presented in Table\,\ref{tab:Comparative_NAS09}.

The differences in the Compton-Y parameter affect the particle distribution and ultimately the observed SSC spectrum. In particular, particles are able to cool below $\gamma_c$ in our model due to the additional cooling by adiabatic expansion. This introduces an additional cooled distribution of electrons that scales as $N_e(\gamma)\propto \gamma^2$, which is similar to a thermal distribution that should form at mildly relativistic and non-relativistic particle velocities.

Fig.\,\ref{FIG_COMPARISON_NAKAR} compares the smoother SSC spectrum obtained from our model to the sharply broken power-law approximation of NAS09. To normalize the spectrum from NAS09, we use the flux normalization from \citet{Sari+98}, which always shows a larger overall normalization compared to our results (also see Fig.\,\ref{fig:EATS_LOS_COMPARISON}). We find that the spectral shapes of both the KN modified synchrotron spectrum and the IC spectrum generally agree with those obtained from NAS09. However, the locations of the spectral breaks do not match between the two results. 
In particular, when comparing the EATS integrated spectrum (thick black curve) to that obtained along the LOS (thin black curve), the spectral breaks at $\nu_m$ and $2\nu_m\gamma_m^2$ do not align, where the EATS spectrum shows these break at energies typically larger by a factor of $\sim8$.

\section{MCMC Model Fits to Observational data} \label{sec::MCMC_FITS}

To fit our model to observational data, we perform Bayesian inference of our model parameters that describe the afterglow lightcurve and spectrum. We implemented a Markov Chain Monte Carlo (MCMC) sampler employing the open-source Python package \texttt{emcee}  \citep{Foreman-Mackey-et-al-2013}. This method is based on the affine-invariant ensemble sampler proposed by \citep{Goodman-Wear-2010} which is particularly efficient for exploring complex, high-dimensional parameter spaces that involve correlated variables, such as those that described the GRB afterglow emission. Our likelihood function evaluates the agreement between the fluxes predicted by our SSC model and the observed data at different times and frequencies. Then the total log-likelihood function is given by
\begin{equation}
\log \mathcal{L} = -\frac{1}{2} \sum_j\sum_i \left[ \frac{ \left([\nu {F_\nu}]_{j,i}^{\rm obs} - [\nu {F_\nu}]_{j,i}^{\rm model} \right)^2}{{\sigma_{j,i}}^2} + \ln(2\pi {\sigma_{j,i}}^2) \right],
\end{equation}
where \([\nu {F_\nu}]_{j,i}^{\rm obs}\), \([\nu {F_\nu}]_{j,i}^{\rm model}\), and \(\sigma_{j,i}\) are the observed flux, the model-predicted flux, and the flux uncertainty at the \(i\)-th energy data point and \(j\)-th observed time, respectively. The posterior probability is constructed by combining the likelihood with uniform priors over physically motivated ranges of the model parameters. 
Our model is described by \(N_{\rm dim} = 7\) parameters, and we use \(N_{\rm walkers} = 21\) (i.e., three times the dimensionality). Chains are initialized around a reasonable set of values of physical parameters with small random perturbations. We further assume that the priors are uniformly distributed within physically motivated bounds. In appendix \ref{sec:MCMC_synthetic_data}, we verify our MCMC fitting routine by creating and fitting to synthetic afterglow data.

\subsection{MCMC Model Fits to GRB 190114C}
We now apply the MCMC method to fit our model to the multi-waveband afterglow of GRB\,190114C \citep{MAGIC_GRB190114C}, which is the first GRB with a TeV afterglow \citep{MAGIC-GRB190114C}. The prompt emission had a duration of $T_{\rm GRB}\approx25$\,s and an isotropic-equivalent $\gamma$-ray energy of $E_{\gamma,\rm iso}=(2.5\pm0.1)\times10^{53}$\,erg at a redshift of $z=0.4245$. 

%
%
For the fit we consider the afterglow spectrum at three different times, where the actual spectrum at any given time is an average over some duration $\Delta T = T_f-T_i$. Since GRBs are photon starved at high energies, and especially at TeV energies, observations have to be binned over time to obtain a reasonable signal-to-noise ratio. When fitting the data with our model, it becomes computationally intensive to first average the spectrum over some interval $\Delta T$. Instead, we take a flux weighted mean time,
\[
\langle T \rangle = \left( \frac{1 - \alpha}{2 - \alpha} \right) \cdot \frac{T_f^{2 - \alpha} - T_i^{2 - \alpha}}{T_f^{1 - \alpha} - T_i^{1 - \alpha}}\,,
\]
to calculate the model spectrum, where we assume that $F_\nu \propto T^\alpha$ and $\alpha \neq 2$. The power-law index $\alpha$ is obtained from the observed lightcurve. For a reliable estimate we use the X-ray (1-10 keV) lightcurve, for which $\alpha_{\rm X}=-1.36 \pm 0.02$ \citep{MAGIC_GRB190114C}. In addition, we impose a constraint on the deceleration time of the blast wave, where we set  
$T_{\rm dec}=20 \, \rm sec$ which is the approximate timescale beyond which the X-ray and high-energy emission shows a smoothly declining lightcurve.

\begin{figure*}
    \centering
    \includegraphics[width=0.9\linewidth]{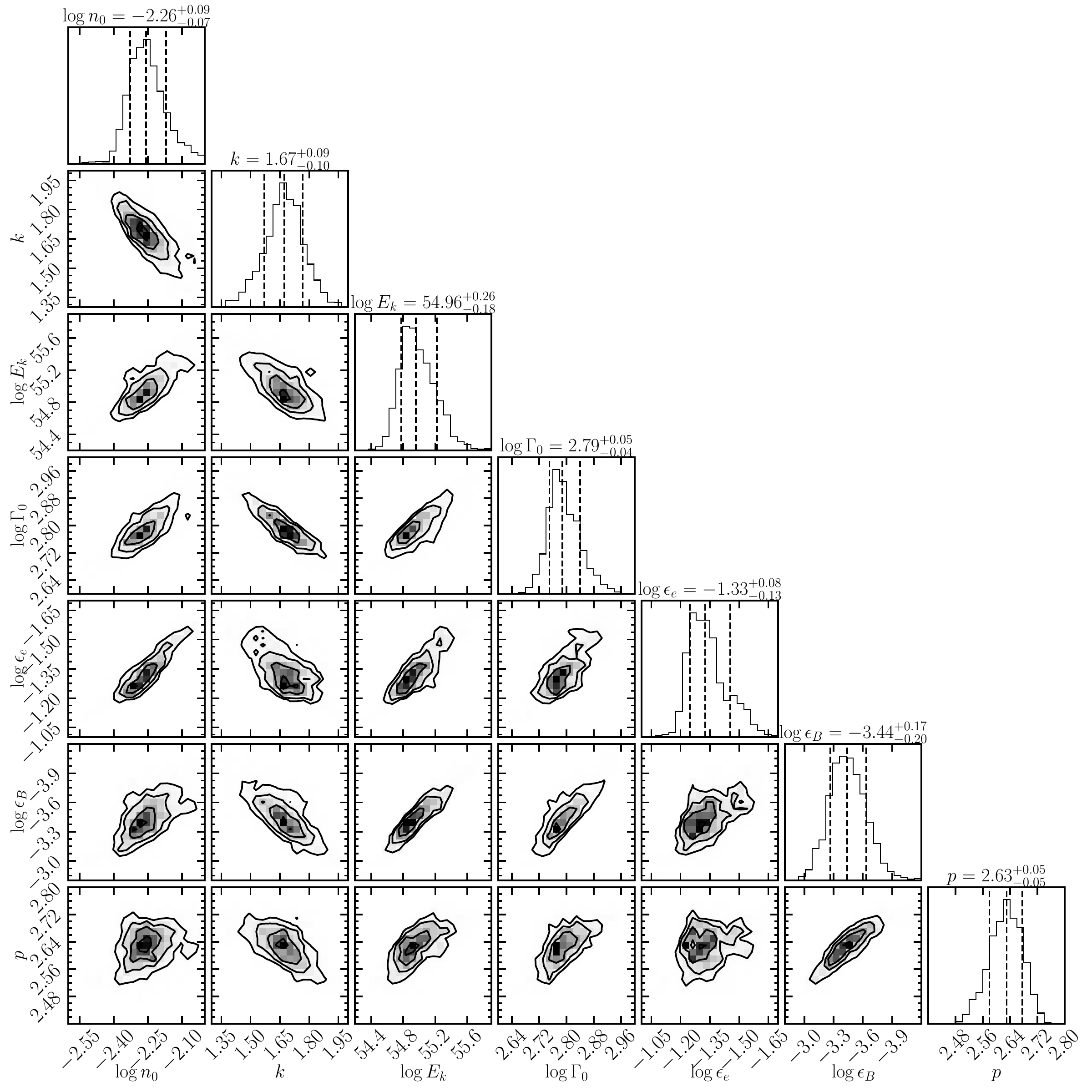}
    \caption{ Posterior distributions of the model parameters inferred from an MCMC fit to the broadband afterglow observations of GRB 190114C. The diagonal panels show the marginalized distribution for each parameter and the dashed vertical lines indicate the median value and the $1\sigma$ confidence intervals.}
    \label{fig_CORNER_data}
\end{figure*}

\begin{figure*}
\centering
\begin{minipage}[b]{0.50\linewidth}
\centering
\includegraphics[width=1.\linewidth]{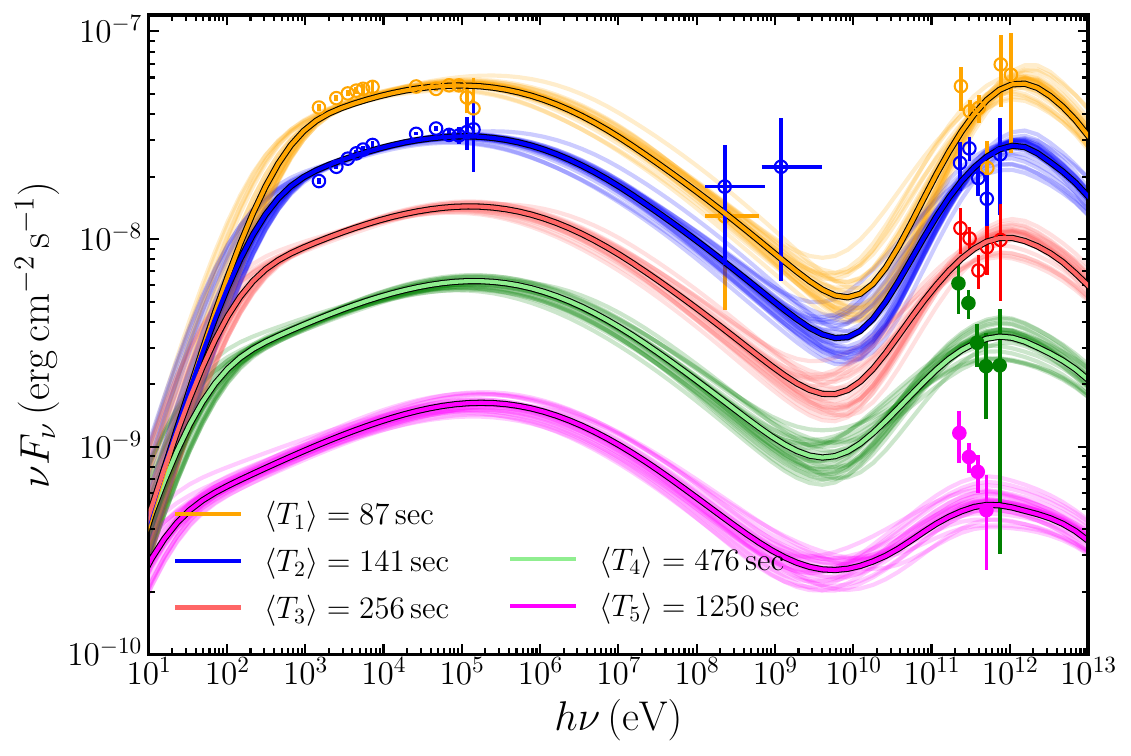} 
\end{minipage}\hfill 
\begin{minipage}[b]{0.50\linewidth}
\centering
\includegraphics[width=1.\linewidth]{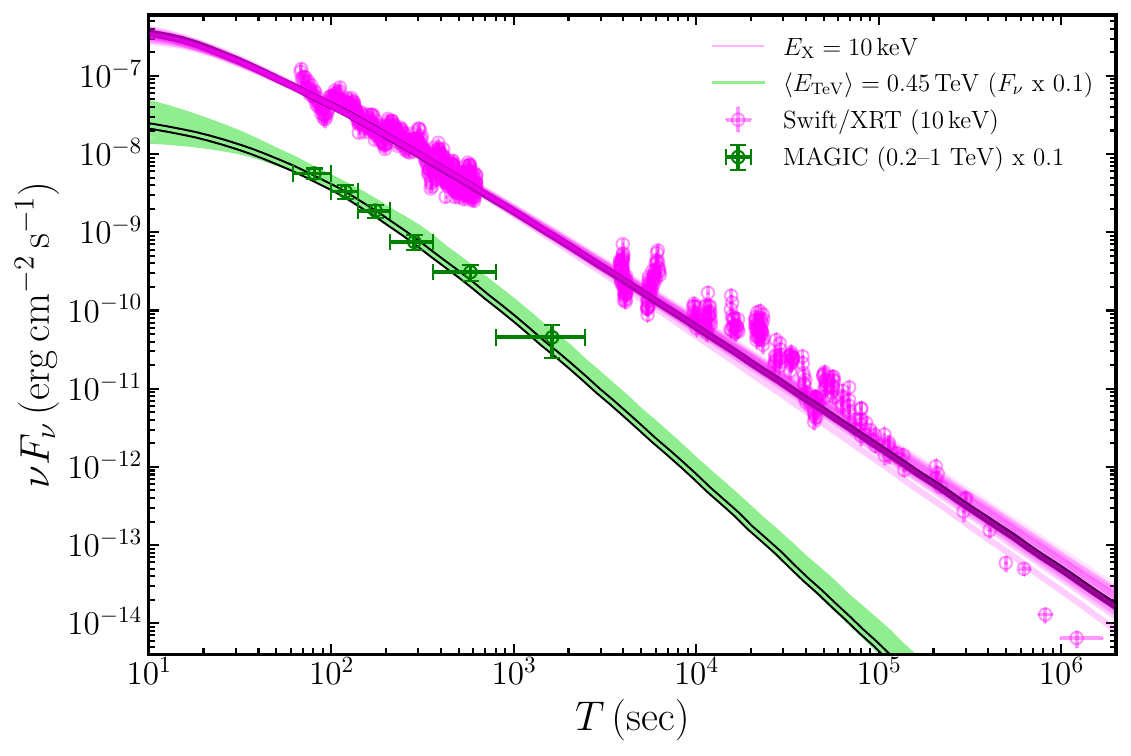} 
\end{minipage} 
\vspace{-16pt} 
\caption{
The spectra and lightcurves corresponding the best-fit inferred parameters (bold curves) and others randomly sampled from the model parameter posterior distributions (light curves). The inferred parameters are listed in Table \ref{tab:Comparative_GRB190114c}.  Thick solid lines represent the best-fit values and thin transparent lines represent solutions within the $1\sigma$ of confidence levels. 
\textbf{Left)} Spectra for three time intervals $\langle T_1 \rangle = [68-110] \, \rm sec$, $\langle T_2 \rangle = [110-180] \, \rm sec$ and $\langle T_3 \rangle = [180-360] \, \rm sec$. \textbf{Right)} Lightcurve were calculated for X-rays at $E_{\rm X} = 10 \, \rm keV$ and midpoint energy for the MAGIC data $\langle E_{\rm TeV} \rangle = [0.2-1] \, \rm TeV$.}
\label{fig_FIT_SPECTRA_LC}
\end{figure*}

\begin{table*}
\centering
\setlength{\tabcolsep}{6pt} 
\renewcommand{\arraystretch}{1.2} 
\begin{tabular}{c c c c c c c c c c}
\multicolumn{10}{c}{\textbf{GRB 190114C}} \\
\midrule
  $T \, [\rm sec]$ & 
  $ E_{k, \rm iso} \, [10^{54} \, \rm erg] $ & 
  $ \Gamma_0 $ & 
  $ n_{\text{0}} [\rm cm^{-3}]  (A_*) $& 
  $ k $ &     
  $ \epsilon_e  \, [10^{-2}] $ & 
  $ \epsilon_B  \, [10^{-4}] $ &  
  $ p $ &  
  $ \xi_e $ &
   \text{Refs.}                %
                   \\
\midrule
 $87$, $141$, $256$ & $9.1^{+7.41}_{-3.13}$  & $613.2^{+78.7}_{-50.1}$ & $5.52^{+1.25}_{-0.82} \times 10^{-3}$ & $1.67^{+0.09}_{-0.10}$ & $4.73^{+0.94}_{-1.22}$ & $3.65^{+0.18}_{-1.33}$ & $2.63^{+0.05}_{-0.05}$ & 1 & \textbf{This work}  \\   
  & &  & ($1.87 \times 10^{-2}$) &  &  &  & &  &   \\   
    $[60-110]$, $[110-180]$& 0.8  & - & 0.5 & 0 & 7 & 0.8 & 2.6 & 1 & M19$^{\dag}$\\   
    100 & 0.6  & 300 & $300$ & 0 & 7 & 40 & 2.5 & - & W19$^{\dag}$  \\
    80& 1  & 600 (300) & 1.0 (0.1)& 0 (2) & 6 (8)& 9 (12) & 2.3(2.35) & 0.3 & A20$^{\dag}$  \\    
    90, 145& 0.3  & - & 2.0 & 0 & $\sim$10 & $\sim$27-61 & 2.5 & $\neq 1$ & DP21$^\dag$ \\
    90, 145& 0.3  & - & ($4.6 \times ^{-2}$) & 2 & $\sim$11-13 & $\sim$ 26-62 & 2.5 & $\neq 1$ & DP21$^\dag$ \\
    90, 150& 4  & - & ($2 \times10^{-2}$) & 2 & 3.3 & 120 & 2.18 & 1 & JR21$^{\dag}$  \\
   $[66-92]$
   & 2.0  & - & $1.06$ ($6 \times 10^{-2}$) & 0 (2) & 1 & $5\times10^{-2}$ & 2.3 & 1 & F19$^\dag$  \\
   $[68-110]$
   & 0.63  & 500 & $0.2$ & 0 & 5 & 50 & 2.8 & - & FT25$^{\dag}$  \\
   $[60-110]$, $[110-180]$ & 10  & 600 & 0.5 & 0 & 1 & 0.1 & 2.6 & 1 & N25$^{\dag}$\\   
\midrule

\end{tabular}
\caption{Comparison of our best-fit model parameters with those obtained in different works.  
\\
\textbf{Notes:} M19: \citep{MAGIC_GRB190114C}, W19: \citep{Wang-et-al-19}, A20: \citep{Asano-Murase-Kenji-20}, DP21: \citep{Derishev-Piran-21}, JR21: \citep{Joshi-Razzaque-2021}, F19: \citep{Fraija-et-al-19}, FT25: \citep{Foffano-Tavani-2025}, N25: \citep{Nedora+25}. We indicate with a $\dag$ those works that either fix the values of some model parameters a priori and/or manually find the best-fit parameters and do not use MCMC. In this work we find $k<2$, but we quote an $A_*$ value when assuming $k=2$ for comparison with other works.
}

\label{tab:Comparative_GRB190114c}
\end{table*}

Figure\,\ref{fig_CORNER_data} shows the posterior distributions of the model parameters obtained from the MCMC fit. We find that all of the model parameter are well constrained, with relatively narrow spreads around their median values. Furthermore, they all exhibit some degree of correlations that reflects degeneracies between the different model parameters due to insufficient constraints from observations. 

The distribution of the kinetic energy of the blast wave shows a peak around $E_{\rm k,iso}\simeq 9.1\times10^{54}$\,erg.
When compared with the prompt $\gamma$-ray energy release of $E_{\gamma,\rm iso}\simeq2.5\times10^{53} \, \rm erg$, this yields a low efficiency of $\eta_{\gamma} = E_{\gamma,\rm iso}/ ( E_{\rm k, iso} + E_{\gamma,\rm iso} ) \approx 2.7\%$ when compared to other TeV non-bright bursts that typically show $\eta_\gamma\sim15\%$ when $\epsilon_B\sim10^{-4}$ \citep{Beniamini+16}. 

In most works fitting the multi-waveband afterglow observations, including the TeV data, a strict assumption for $k=0$ or $k=2$ is made as a simplification. In our approach the radial profile of the external medium density is free to vary during the fit and is constrained by observations. Here we find that the external medium has a wind-like radial density profile with $k \approx 1.7$, where $k=2$ is expected in the collapsar scenario of long-duration GRBs under the simplifying assumptions of a constant mass outflow from the progenitor star, $\dot M = 4\pi r^2 \rho v_w= 10^{-5}M_\odot\,{\rm yr}^{-1}$, at a fixed wind speed of $v_w = 10^3\,{\rm km\,s}^{-1}$. A power-law index of $k<2$  could imply that the medium is not entirely dominated by a steady stellar wind, but rather by a combination of a wind plus a uniform medium, or possibly by another kind of wind mass-loss rate evolution \citep[e.g.,][]{Curra+09, Yi-Wu-Dai-13}.

In Fig. \,\ref{fig_FIT_SPECTRA_LC} we show the best-fit (bold curve) spectra and lightcurves, along with the same calculated by randomly sampling the posterior distributions to show their spread. Our model provides a good fit to the overall data. 
The 10\,keV lightcurve is well explained by the model up to $T\sim (3-4) \times 10^{5} \, \rm sec$, above which the observations show a steepening in the lightcurve, which we interpret as occurring due the jet break. Our model uses a spherical flow and it is designed to explain the early afterglow observations when we expect to observe any TeV emission. This typically occurs before the incidence of the jet break at $0.1\,{\rm d} \lesssim T_j\lesssim 5\,{\rm d}$. Therefore, the steepening in the lightcurve due to this effect is not captured in our model. In the TeV band, our solution exhibits greater dispersion, which is due to the large error bars in the MAGIC data. A similar trend is observed in the Fermi-LAT band. However, the model TeV lightcurve shows an excellent agreement with observations. The TeV spectra at later times, particularly at $\langle T_4 \rangle$ and $\langle T_5 \rangle$, show a steeper spectral trend that is not fully reproduced by our model. As obtained in \citet{Derishev-Piran-21}, it appears to require the SSC peak to be at energies lower than what we find.

We estimate the jet break from the steepening in the X-ray lightcurve that occurs at approximately $T_{j} \sim 3\times 10^{5} \, \rm sec \, (3.47 \, \rm day)$. Using the relationship given by $\theta_j = (T_j/T_{\rm dec})^{\frac{3-k}{2(4-k)}} \Gamma_0^{-1}$ with $T_{\rm dec} = (1+z)r_{\rm dec}/(2c\Gamma_0^2)$ and using the inferred parameters, we obtain a jet open angle of $\theta_j \approx 0.0234 \, \rm rad \, (1.34^\circ)$,  which implies a corrected kinetic energy $E_{j} = \theta_j^2 E_{k, \rm iso}/2 \approx 2.5 \times 10^{51} \,\rm erg$.

\subsection{Comparison with Other Works}

The best-fit parameters found in this work are compared with different works in Table\,\ref{tab:Comparative_GRB190114c}. It is important to mention that while we simultaneously fit spectra at five different times, many of these listed works derive best-fit parameters by fitting only a single spectrum. Here we briefly comment on works that used numerical models whose results may be comparable to our semi-analytical model. 

\cite{Derishev-Piran-21} performed independent fits to spectra at two different times with uncoupled model parameters, resulting in two distinct parameter sets that included some parameters whose value was fixed a priori. They obtain electron (and positron) distribution and photon spectrum in the comoving frame by solving the kinetic equations and, like in this work, account for particle adiabatic cooling and density dilution for both photons and particles. In addition, they include the effects of pair-production due to $\gamma\gamma$-annihilation and include the radiation from these secondary pairs. The comoving radiation field is not EATS integrated, but obtained in the observer frame using EATS averaged \textit{effective} coefficients that relate comoving quantities to that in the observer frame. This amounts to scaling the instantaneous comoving spectrum using these coefficients to obtain the observed spectrum. As shown in Fig.\,\ref{FIG_COMPARISON_NAKAR}, the LOS spectrum is similar in spectral shape to the EATs integrated one, and therefore it is possible to obtain an approximate observer-frame spectrum by appropriately scaling the instantaneous comoving spectrum.

\citet{Asano-Murase-Kenji-20} used an entirely time-dependent kinetic code (their Method I) from \citet{Fukushima+17}, similar to the one used in this work to verify our semi-analytic approach. They also obtained the observer-frame spectrum by integrating the comoving spectrum over the EATS. Therefore, we expect the results of our work to be broadly similar to those from \citet{Asano-Murase-Kenji-20}. However, there are some differences between our approach and theirs. For example, they fix the fraction of shocked electrons that are accelerated into a power-law energy dissipation and whose radiation we observe as SSC to $\xi_e=0.3$ a priori. This also means that the remaining $1-\xi_e=0.7$ fraction of total shocked electrons must emit thermal radiation, which has never been observed in the afterglow phase. For that reason, we make the standard assumption of $\xi_e=1$. As pointed out by \citet{Eichler-Waxman-05}, having this fraction in the range $m_e/m_p \leq \xi_e \leq 1$ results in a degeneracy where the same afterglow spectrum is obtained when $E_{\rm k,iso}=E_{\rm k,iso}^*/\xi_e$, $n_0 = n_0^*/\xi_e$, $\epsilon_e = \xi_e\epsilon_e^*$, and $\epsilon_B=\xi_e\epsilon_B^*$, where the starred parameters are obtained when $\xi_e=1$.

Finally, in a recent work, \citet{Nedora+25} have developed a comprehensive kinetic code that includes reverse shock dynamics and emission as well as jets with angular structure. Their formulation of the SSC emission is broadly similar to ours. When fitting to the broadband spectrum of GRB\,190114C using a spherical shell, they perform the fit over a very coarse grid of model parameters and fix the circumburst environment to an ISM ($k=0$).

As can be seen from Table\,\ref{tab:Comparative_GRB190114c}, a variety of model parameters, some with large dispersion, have been obtained in different works that employ a variety of methods to produce the SSC spectrum. In most works the model parameters have been obtained by comparing the model with the data by eye, and therefore, model parameter degeneracies are not so clear. In addition, many works assume an ISM circumburst medium that makes comparison with our results difficult due to the inherent degeneracies. 
Even when comparing our results to only those models that assume a stellar wind medium ($k=2$) and use kinetic approaches \citep[e.g.][]{Asano-Murase-Kenji-20, Derishev-Piran-21}, 
there is no strong consensus.

An exception are \cite{MAGIC_GRB190114C} and \citep{Nedora+25}, they fit two observed time intervals simultaneously, however, their analysis assumes an ISM ($k=0$), and naturally their inferred model parameters differ significantly from ours, particularly in the larger value of the ratio $\epsilon_e/\epsilon_B \sim 875-1000$. In general the rest of models listed, for ISM medium, predicts $\epsilon_e/\epsilon_B \sim 10-10000$ ratios and lower $E_{k, \rm iso}$ respect to our results.

On the other hand, when comparing our results only with those models that assumes a stellar wind medium, $k=2$, and use kinetic approaches \citep[e.g.][]{Asano-Murase-Kenji-20, Derishev-Piran-21}, our results are closer with their findings with some deviations. For example, our finding indicates $\epsilon_e/\epsilon_B \sim 126$ and $E_{k, \rm iso} \approx 9.1 \times 10^{54} \, \rm erg $, this is in contrast to the assumed lower value $E_{k, \rm iso} \sim 3 \times 10^{53} \, \rm erg $ and $\epsilon_e/\epsilon_B \sim 17-50$ by \cite{Derishev-Piran-21}. Similar, \cite{Asano-Murase-Kenji-20} assume $E_{k, \rm iso} \sim 1 \times 10^{54} \, \rm erg $ requiring $\epsilon_e/\epsilon_B \sim 67$ and $\Gamma_0 = 300$ which are lower values respect to our finding. Such discrepancies among different works even for the same circumburst environment and similar numerical treatment may arise due to parameter degeneracies as indicated by our MCMC sampling. In terms of $E_{k,\rm iso}$, $\Gamma_0$, $p$ our results are similar to \citep{Nedora+25} but they assume $k=0$ and find $\epsilon_e/\epsilon_B = 1000$.

\section{Summary \& Discussion}\label{sec::CONCLUSIONS}

Our findings demonstrate that our proposed semi-analytic approach effectively captures the temporal evolution of the SED, before and after the deceleration radius, with results very similar to that obtained using a numerical kinetic code (see Fig.\,\ref{fig:Comparison}). Furthermore, we demonstrate that neglecting the effect of adiabatic expansion and adopting different radiation escape timescales significantly modify the resulting spectra. In particular, this omission causes a substantial overestimation of the SSC flux component and modifies the spectral shape due to an incorrect prediction of the characteristic frequencies in both the synchrotron and the SSC components, especially in the slow-cooling regime (see Fig.\,\ref{fig_spectra_effects}). 

When we compared our model with the analytical formalism of NAS09, we found significant discrepancies between the two. These arise due to some simplifications in the NAS09 formalism that ignores the effects of adiabatic cooling, dilution of the radiation field due to expansion, a finite rate of photon escape, and EATS integration. Furthermore, we find that there is no configuration within our framework that is fully equivalent to NAS09, because they assume that all SSC photons escape instantaneously as they are produced at some radius $r$. In contrast, escaping photons take a finite amount of time (on the order of the light crossing time of the shell) in our model, and hence only a fraction of all photons produced at radius $r$ are able to escape. As a consequence, we find that the NAS09 approach tends to overestimate the actual Compton-Y parameter and the SSC flux (see Fig.\,\ref{fig_Y_ne_comparison} \& \ref{FIG_COMPARISON_NAKAR}), which has implications for the correct estimation of the physical parameter values when it is applied to fit observational data. 

Moreover, spectral shape discrepancies between NAS09 and our model are not solely due to the aforementioned effects. They also arise from an oversimplified approximation of the scattering kernel that smoothly transitions from the Thomson to the KN regime over three orders in magnitude of the argument of the kernel, given by the variable $\widetilde{x}_{\rm KN}$ (see Fig.\,\ref{fig:fKN}). We showed that the widely used step function approximation cannot properly capture the actual behavior of the Compton-Y parameter (see Fig.\,\ref{fig_Y_ne_comparison}).
It makes the spectrum in NAS09 very sensitive to the value of $\widetilde{x}_{\rm KN}$ at which an abrupt transition from the Thomson to the KN regime is made, as shown in Fig.\,\ref{fig_SSC_xKN_data}. 
This artifact can be remedied by adopting better approximations of the scattering kernel that may offer more accurate results while keeping the treatment analytic (Aguilar-Ruiz et al, in preparation).

We applied our formalism using MCMC to fit the observed afterglow spectra and lightcurve of GRB\,190114C. 
We inferred a value of $E_{k, \rm iso} \approx 9.1 \times 10^{54} \, \rm erg$ for the isotropic-equivalent kinetic energy, which suggests that this GRB is more energetic than what was previously shown by other works that assume $k=2$ and find $E_{k,\rm iso}\lesssim 4 \times 10^{54} \, \rm erg$.
Such high $E_{k,\rm iso}$ in this GRB puts it closer to GRB\,221009A, a.k.a the BOAT, which was the brightest of all time GRBs that had a kinetic energy budget of $E_{k,\rm iso}\sim2\times10^{55}$\,erg for a jet with a shallow angular profile \citep{OConnor+23,Gill-Granot-23}. It was also the brightest in the TeV band with more than 64,000 photons detected at $E>0.2$\,TeV \citep{LHAASO+23}. In light of this, our results suggest that afterglow TeV emitters require large $E_{k,\rm iso}$ energy budgets to power the TeV emission and that they may represent the extreme class of GRBs.
In addition, these GRBs also tend to exhibit high prompt $\gamma$-ray fluences that form a rare subset around 1 per cent of all GRBs \citep{Noda-Parsons-22}.

Our modeling self-consistently finds a value of $k= 1.67$ for the power-law index of the external medium density, which supports its wind-like nature that is expected in collapsar driven GRBs over an ISM.
Since we find $k < 2$, it also opens up the possibility that
the external medium is not entirely dominated by a constant stellar wind, 
and may support either a non-uniform mass outflow rate from the progenitor star or
a combination of a stellar-wind and an ISM \citep[e.g.][]{Fraija-et-al-19}, with the former transitioning into the latter at large distances from the central engine. Finding a value for $k$ different from the generally assumed fixed values of $k=0$ and $k=2$ also underscores the need to treat it as a free parameter when comparing with observations. Since afterglow parameters are generally degenerate, fixing the value of $k$, although convenient, may lead to an inaccurate inference of the other model parameters. This is supported by the posterior distributions that we obtain from our MCMC fit that shows correlations between the external medium properties and the jet dynamical parameters. 

One of the largest uncertainties in the physics of afterglow shocks is the fraction of energy in the shocked fluid that goes into the radiating electrons and magnetic field. Here we find both values, with $\epsilon_e \simeq 4.6 \times 10^{-2}$ and $\epsilon_B \simeq 3.6 \times 10^{-4}$, that agree with canonical expectations. In the Thomson regime, the value of the Compton-Y parameter for a slow-cooling spectrum is given by $Y_{\rm Th} = (\nu_c/\nu_m)^{(2-p)/2}(\epsilon_e/\epsilon_B)^{1/2}\simeq2$, where we have used the value of $p=2.63$ from our fits and the critical frequencies from the first two spectra shown in Fig.\,\ref{fig_FIT_SPECTRA_LC}. This estimate agrees with the spectra of GRB\,190114C at those times. The X-ray data in this GRB does not provide strong constraints on the ratio $\nu_c/\nu_m$, which can be used to constrain the ratio of $\epsilon_e/\epsilon_B$ for a given $Y_{\rm Th} = (\nu F_\nu)_{\rm IC}/(\nu F_\nu)_{\rm syn}$ when KN effects are absent. This result is also strictly valid when adiabatic dilution is ignored, which otherwise produces a lower $Y_{\rm Th}$. 

The microphysical parameters $\epsilon_e$ and $\epsilon_B$ are correlated, and our results indicate a ratio of $\epsilon_e / \epsilon_B \sim 127$. This ratio is generally higher by a factor of $\sim 2-5$ to those found by other numerical works that assume $k = 2$. However, comparing with analytical results of \cite{Joshi-Razzaque-2021} who report $\epsilon_e / \epsilon_B \sim 2.75$, our results are almost two orders of magnitude higher. These differences may largely arise due to the missing effects in their analytic treatment, namely adiabatic cooling of particles and dilution of the radiation field.     

Our current formalism only explores the afterglow emission from a spherical blast wave. GRB jets have been shown to have angular structure \citep[e.g.][]{Gill-Granot-18b,Gill-Granot-23,BGG2020} which is a natural outcome of their interaction with the confining medium, i.e. the stellar material in the case of collapsar produced GRBs \citep[e.g.][]{Gottlieb+21}. Therefore, it becomes important to properly account for the jet structure, which was shown to be critical in explaining the afterglow of the TeV bright GRB\,221009A \citep{OConnor+23}. Some attempts have been made to include it in calculating the SSC emission by \citet{Hope+25}, who used a kinetic code and discretized the energy structure of the jet into several uniform zones while keeping the same initial bulk Lorentz factor in all. Kinetic codes are typically very computationally expensive and do not allow performing MCMC fits over a vast parameter space, which becomes even more prohibitive for a jet with angular structure both in energy and bulk Lorentz factor. Since our current formalism is semi-analytic, it easily allows for incorporating more complex jet structures while maintaining acceptable computation times. This will be a topic of a future work (Aguilar-Ruiz et al in preparation).

The number of TeV bright GRBs have grown after the discovery of GRB\,190114C, which now makes it possible to better constrain the afterglow shock physics by including the additional constraints provided by the TeV emission. Future discoveries will indeed assist in answering the question of what really makes GRBs TeV bright.

\section*{Acknowledgements}
PB's work was funded by grants (no. 2020747 and no. 2024788) from the United States-Israel Binational Science Foundation (BSF), Jerusalem, Israel and by a grant (no. 1649/23) from the Israel Science Foundation.
\section*{Data Availability}
No new data was generated or analyzed in this work. 



\bibliographystyle{mnras}
\bibliography{refs} 



\appendix

\section{Compton parameter derivation}\label{Appendix:Compton_Param_derivation}
\subsection{Compton's energy loss rate}
The interaction rate of a single electron passing through a photon radiation field, in the electron's rest frame,  is given by the expression
\begin{equation}
\frac{dN''}{dt''} = \int c \, n_E''(E'',\Omega'') dE'' d\Omega'' d\sigma''_{\rm KN} ,
\end{equation}
where $n_E''(E'', \Omega'') dE'' d\Omega''$ and $d\sigma''_{\rm KN}$ are the differential photon distribution of the radiation field and the full differential cross-section for Compton scattering which includes KN effects, respectively.
Using invariant quantities, $\frac{n_E''(E'', \Omega'') dE'' d\Omega''}{E''}= \frac{n_E'(E', \Omega') dE' d\Omega'}{E'}$, $N_e'' =  N_e'$ and $d\sigma''_{\rm KN} = d\sigma_{\rm KN}'$, the interaction rate is transformed to the comoving fluid frame as 
\begin{align}
\frac{dN_e}{dt'}	&= \frac{1}{\gamma}\frac{dN_e}{dt''} 
	\\
	&= c \int  dE' d\Omega' \, n_E'(E', \Omega') \, (1-\beta_e \mu')  \, dE'_{\rm IC} d\Omega'_{\rm IC} \, \frac{d\sigma_{\rm C}}{d\epsilon'_{\rm IC} d\Omega'_{\rm IC}} \, ,\label{eq_int_rate}
\end{align}
where $\beta_e = \sqrt{1 - 1/\gamma^2}$ is the velocity of the colliding electron in units of the speed of light and $\mu =  \cos \theta_{\gamma e}$ is the collision angle between the electron and the incident photon.
The energy loss rate of a single electron is given by the expression
\begin{align}
P'_{\rm IC} (\gamma) 
&
= \frac{dN_e}{dt'} \left( E'_{\rm IC} - E' \right) \, ,
	 \\
    &= c \int  dE' d\Omega' \,  n_E'(E', \Omega') \, (1-\beta_e \mu') \, \left[ \left\langle E_{\rm IC} \right\rangle - E' \right] \sigma_{\rm KN} (k'') \, ,
 \label{eq_int_rate}
\end{align}
where $\left\langle E'_{\rm IC} \right\rangle
    = \frac{1}{\sigma_{\rm KN}} \int dE'_{\rm IC} d\Omega'_{\rm IC} \, E'_{\rm IC}
 \, \frac{d\sigma'_{\rm KN}}{dE'_{\rm IC} d\Omega'_{\rm IC}}
 $ is the mean energy of scattered photons. After performing the integration, this expression becomes \citep[see][]{Dermer-Menon-09}
 %
 %
\begin{multline}\label{eq_Eic_average}
    \left\langle E'_{\rm IC} \right\rangle 
    = 
    \gamma \, m_e c^2 \, \Bigg\{ 1 - \frac{3\sigma_{\rm T}}{8{{k''}^3 \sigma_{\rm KN}}}
    \Bigg[ \frac{{k''}^2}{3} \left( \frac{(1+ 2k'')^3-1}{(1+2k'')^3} \right) +
    \\
    + \frac{2k'' \left({k''}^2 -k'' -1 \right)}{(1+2k'')} + \log(1+2k'') \Bigg] \Bigg\} \, ,
\end{multline}
where $k'' = \gamma (1- \beta_e' \mu') \, E / (m_e c^2)$ is photon energy in the electron's rest frame normalized by the electron rest-mass energy. Similarly, the total Compton cross-section is defined as $\sigma_{\rm KN} = \int dE'_{\rm IC} d\Omega'_{\rm IC} 
 \, \frac{d\sigma'_{\rm KN}}{dE'_{\rm IC} d\Omega'_{\rm IC}}$ and after integration it takes the following expression \citep[see][]{Dermer-Menon-09, Rybicki-Lightman-79}
 
%
%
\begin{align}\label{eq_Compton_cross_section}
    \sigma_{\rm KN}
    =  \frac{3}{8} \frac{\sigma_{\rm T}}{{k''}^2} \left[ 4 + \frac{2 {{k''}}^2 (1+ {k''})}{(1+2{k''})^2} + \frac{{{k''}}^2 -2{k''} -2}{{k''}} \log(1+2{k''}) \right] \,.
\end{align}

In the case of an isotropic photon distribution and ultra-relativistic electrons, $\gamma\gg1$, the energy's 
loss rate of an electron crossing an  isotropic radiation field is given by
\begin{align}\label{eq_Compton_energy_loss}
P'_{\rm IC}(\gamma)
&= 
	c \int_0^\infty  dE' \, n_E'(E') \int_{-1}^{+1}
 d\mu \frac{(1-\mu)}{2}   \, \left[ \left\langle E'_{\rm IC} \right\rangle - E' \right] \sigma_{\rm KN} ( k'' ) \, .
\end{align}

%
%
%

\subsection{Compton-Y Parameter for an isotropic photon radiation field and ultra-relativistic electrons}

%
%
The Compton-Y parameter, $Y_{\rm ssc}$, is defined by Eq. \ref{eq_Yssc_PARAM_DEF} as the ratio of energy losses of Compton scattering and synchrotron emission, both produced by the same electron population. Taking Eq. \ref{eq_Compton_energy_loss} for Compton energy losses and synchrotron energy losses, given by
\begin{equation}
    P'_{\rm syn} (\gamma) = \frac{4}{3} \sigma_{\rm T} c \, u_B' \gamma^2 \beta_e^2\, .
\end{equation}
in the case of ultra-relativistic electrons, $\beta\approx1$, hence, the Compton-Y parameter takes the expression
\begin{align}
    Y_{\rm ssc}(\gamma) &= \frac{P'_{\rm IC}(\gamma)}{P'_{\rm syn} (\gamma)}
    \\
    \\
    &= 
     \int_0^\infty  d\nu' \, \frac{u'_{\nu'}}{ u_B' } \int_{-1}^{+1}
 d\mu \frac{3(1-\mu)}{8 h \nu' {\gamma}^2 }   \, \left[ \left\langle E_{\rm IC}' \right\rangle - h \nu' \right]  \frac{\sigma_{\rm KN} ({k''} )}{\sigma_{\rm T}} \, ,
    \end{align}

Expressing the above equation in terms of synchrotron photon frequencies $\nu'$ and $ \widetilde{\nu}'$ (see Eq. \ref{eq_nu_tilde}) and using the differential energy density $u'_{\nu'} = h \nu' \, n'_{\nu'}$, the general expresion of Compton-Y parameter for relativistic electrons takes the form
\begin{equation}
    Y_{\rm ssc} (\gamma) = \frac{1}{u_B'} \int_0^\infty  d\nu' \, u'_{\nu'} \, \left[ f_{\rm KN_1} \left( \frac{\nu'}{\widetilde{\nu}'} \right) + \frac{1}{\gamma^2} f_{\rm KN_2}\left( \frac{\nu'}{\widetilde{\nu}'}\right) \right]
\end{equation}
where we define the last two quantities as
\begin{multline}
    f_{\rm KN_1} (\widetilde{x})= \frac{3}{8} \frac{1}{\widetilde{x}}\int_{-1}^{+1} 
 d\mu (1-\mu)   \, \frac{\left\langle E_{\rm IC}'\right\rangle}{\gamma m_e c^2} \frac{\sigma_C\Big(\widetilde{x}(1-\mu)\Big) }{\sigma_{\rm T}} \, ,
 \\
 f_{\rm KN_2} (\widetilde{x}) = \frac{3}{8} \int_{-1}^{+1} 
 d\mu (1-\mu)   \, \frac{\sigma_C\Big(\widetilde{x} (1-\mu)\Big) }{\sigma_{\rm T}} \, ,
 \\
\end{multline}
with $\widetilde{x} = \frac{v'}{ \widetilde{\nu}' }$. In general, when $\left\langle E_{\rm IC}' \right\rangle \gg h \nu'$, the second term can be neglected.

\section{Continuity equation}

\subsection{Electron continuity equation}\label{Appendix:electron_cont_eq}

The evolution of particles in time and energy is governed by the continuity equation. Assuming that particles neither escape nor undergo further acceleration, the electron continuity equation in the fluid's comoving frame is written as
%
%
\begin{equation}
    \frac{\partial n'_e(\gamma,t')}{\partial t'} + 
     \frac{\partial}{\partial\gamma} \left[ n'_e(\gamma,t')  \, \dot{\gamma}'_{\rm cool}(\gamma, t') \right] = q'_e(\gamma,t') - \frac{n_e'(\gamma,t')}{t'_{\rm ad}(t')}
\end{equation}
where $\dot{\gamma}'_{\rm cool}$ is the total cooling rate, $t'_{\rm ad}(t')$ is the adiabatic timescale, and $q'_e(\gamma, t')$ is the injection rate of electrons, which in general  is assumed to be
%
%

\begin{equation}\label{eq_e_InjectionDistribution}
\frac{\partial^2n_e'(\gamma,t')}{\partial t'\,\partial\gamma} =  {q'_e}({\gamma}, t') = {q'_0} (t') \, \gamma^{-p} \, 
\qquad \gamma_{m} \leq \gamma \leq \gamma_{M}\,,
\end{equation}

\subsubsection{Steady state solution} \label{Appendix:cont_eq_sol}

If neither cooling nor adiabatic expansion modifies the electron distribution in a small interval of time, i.e. when $\Delta t' \ll \min(t'_{\rm cool}, t'_{\rm ad})$ such that, $n'_e(\gamma)$ remains roughly constant in the time interval $[ t', t'+\Delta t' ]$, the electron distribution can be treated as quasi-steady. In this case, the continuity equation is reduced to a first order differential equation, such that
\begin{equation}
   - \frac{\partial}{\partial\gamma} \left[ n'_e(\gamma,t')  \, |\dot{\gamma}'_{\rm cool}(\gamma, t')| \right] = q'_e(\gamma,t') - \frac{n'_e(\gamma,t')}{t'_{\rm ad}(t')} \, .
\end{equation}
This equation admits two solutions determined by the condition of whether $\gamma>\gamma_m$ or $\gamma<\gamma_m$. In the following we present both cases.

\paragraph*{Solution for $\gamma>\gamma_m$:} 
This equation can be solved, using a standard method for solving first ordinary differential equation, and taking the the boundary condition as $n_e(\gamma_{\rm M})=0$,
the solution is
\begin{equation}\label{eq_eD_SSC_AD}
    n_{\rm e}'(\gamma) = \frac{1}{|\dot{\gamma}'_{\rm cool}  (\gamma)|} \int_{\gamma}^{\gamma_{\rm M}} q'_{e} (\gamma') \, I_e(\gamma',\gamma)\, d\gamma' \, , \quad \gamma_m < \gamma < \gamma_M \, , 
\end{equation}
where the function $I_{\rm e}(\gamma',\gamma)$ is the ratio of the integrating factors, given by
\begin{equation}
    I_e(\gamma',\gamma) = \exp \left[ - {  \int_{\gamma}^{\gamma'} \frac{ d\gamma''}{t'_{\rm ad} \, |\dot{\gamma}'_{\rm cool}    (\gamma'')| } } \right]\, .
\end{equation}
Using integral properties, this factor can be expressed as $I_e(\gamma', \gamma) =\frac{I_e(\gamma',\gamma_M)}{I_e(\gamma, \gamma_M)}$ which is particularly useful for numerical implementation.
The above solution matches to the independent-time electron escape solution \citep[see Eq. C10 in][]{Dermer-Menon-09}, therefore, analogous to that case, the function $I_e(\gamma', \gamma)$ can be interpreted as the probability to an electron with LF, $\gamma'$, cools down to $\gamma$ before it lost energy due to adiabatic expansion.

\paragraph*{Solution for $\gamma<\gamma_m$.} 
In this case the continuity equation also have a solution given by
\begin{multline}
    n'_{\rm e}(\gamma) =\frac{1}{ |\dot{\gamma}'_{\rm cool}  (\gamma)|} \int_{\gamma_m}^{\gamma_{\rm M}} q'_{\rm e} (\gamma') \, I_e(\gamma',\gamma)\, d\gamma' \, , \quad \gamma < \gamma_m \, \\ 
\end{multline}
    where the normalization constant can be determined by connecting both solutions at $\gamma_m$.

%
%

\subsection{Photon continuity equation}\label{Appendix:photon_cont_eq}

The evolution of the photon distribution is described by the continuity equation, where photons 
are added due to SSC emission and removed by the escape term, and the distibution is diluted by 
the adiabatic term, 
\begin{equation}
   \frac{\partial n_{\nu'}'(\nu',t')}{\partial t'} = 
   \frac{ P_{\nu'}'^{\rm ssc}(\nu',t')}{h \nu'} - \frac{n'_{\nu'}(\nu',t')}{t'_{\rm esc}} 
   - \frac{n'_{\nu'}(\nu',t')}{t'_{\rm ad}} \, ,
\end{equation}
where $P_{\nu'}'^{\rm ssc}$ is the SSC radiated power per unit volume and frequency, and $t_{\rm esc}'$ and $t_{\rm ad}'$ are 
the photon escape and adiabatic expansion timescales. 
We solve this equation by discretizing it over a timescale $\Delta t'(r)^{-1}\ll t_{\rm esc}'(r)^{-1} + t'_{\rm ad}(r)^{-1}$, 
and during which it is assumed that the spectral radiated power and the rates of escape and adiabatic dilution 
remain constant. Under these assumptions, the solution can be obtained using the integrating 
factor method, such that
\begin{equation}\label{eq_full_photon_solution}
    n_{\nu'}'(t') 
    =
    \frac{1}{I_\gamma(t', t_0')}\left[ n_{\nu'}'(t'_0) 
    + \frac{{P}_{\nu'}'^{\rm ssc} (t')}{{h \nu'}}\int_{t'_0}^{t'}I_{\gamma}(t', t_0') \, dt' \right]
\end{equation}
where $I_{\gamma}$ is the integrating factor given by
\begin{equation}
    I_{\gamma} (t', t'_0) = \exp \left[ \left( {t'}_{\rm esc}^{-1} + {t'}_{\rm ad}^{-1}  \right) \Delta t'  \right]\,,
\end{equation}
for $\Delta t' = t'-t'_0$. 
The photon distribution has a trivial solution, 
\begin{align}
    n_{\nu'}'(t') 
     &\simeq 
     \exp\left[ - \left( {t'}_{\rm esc}^{-1} + {t'}_{\rm ad}^{-1}  \right) \Delta t' \right] \left[n'_{\nu'}(t'-\Delta t') + \frac{P_{\nu'}'^{\rm ssc}(t')}{h\nu'}\Delta t'\right]\,,  
\end{align}
where the exponential term gives the fraction of the photons that remain in the emission region over time $\Delta t'$. 
The first term inside the second square bracket represents the photon distribution that was injected from the previous 
discretized step, and the second term represents the newly produced SSC photons during time $\Delta t'$. 

Finally, the observed spectral flux will be determined by photons that escape from the emitting region 
after time $\Delta t'$, which escaped spectral power is given by
\begin{equation}
    \frac{P_{\nu', \rm esc}'}{h\nu'} =  \frac{\Delta n_{\nu',\rm esc}'}{\Delta t'} 
    = [1 - \exp(-\Delta t'/t_{\rm esc}')] \frac{\tilde n_{\nu'}'}{\Delta t'}\,,
\end{equation}
where $\tilde n'_{\nu'} = n'_{\nu'}(t')\exp(\Delta t'/t_{\rm esc}')$.

\subsection{Numerical Methodology}\label{Appendix:Num_methods}
The numerical methodology used to solve the continuity equations and compute the observed flux is described as follows:

\begin{enumerate}
    \item \textit{Shell conditions}: For any radius, we start by calculating the physical quantities that define the shell at radius $r$, i.e., $B'$, $\Delta'$, $n_e'$, $\dot{n}_e'$, $t'$, $t'_{\rm ad}$, $t'_{\rm esc}$, and $\gamma_m$.

    \item \textit{Quasi-steady state solution}: For each radius in the range $r_{\rm beam} \leq r \leq r_{\rm LOS}$, the time step is set according to the condition 
    \[
    \Delta t' \ll \min\left( t'_{\rm esc}, t'_{\rm ad}, t'_{e, \rm cool} \right),
    \]
    where $r_{\rm beam}$ is the beaming cone radius and $r_{\rm LOS}$ is the line-of-sight radius. The electron and photon distributions are then obtained iteratively at each radius following these steps:
    
    \begin{enumerate}        
        \item Calculate the Compton parameter, $Y_{\rm ssc}(\gamma, r)$ , using Eq. \ref{eq_Yssc_PARAM}, where the synchrotron photon energy density is given by Eq. \ref{eq_uSSC_sol}, this equation includes the contribution from the previous step, $u_{\nu'}'^{\rm syn}(r-\Delta r)$ and the newly injected synchrotron photons, $j_{\nu'}^{\rm syn}$ given by Eq. \ref{eq_jSYN}.
        \item Calculate the steady-state electron distribution using Eq. \ref{eq_eD_steady_state}, and normalize it using eq. \ref{eq_eD_normalization}.
        \item Update the synchrotron spectral radiated power and the photon energy density using Eq. \ref{eq_jSYN} and Eq. \ref{eq_uSSC_sol}.
        \item If the error in Compton parameter is less than the 5\% finish the iterative method, otherwise go to step (a) and repeat every step.
    \end{enumerate}
    
    \item \textit{Pre-calculation of escaped luminosity}: After convergence is achieved, compute the comoving spectral luminosity of escaping photons,  $L_{\nu'}'^{\rm ssc}(\nu', r)$, using Eq. \ref{eq_Lnu_escape} and store the results.
    \item \textit{Observed flux calculation}: Finally, the observed flux is obtained by integrating the comoving escaped luminosity, $L_{\nu'}'^{\rm ssc}(\nu', r)$, over the EATS for each observer time of interest, using Eq. \ref{eq_EATS_flux}.
\end{enumerate}

\section{MCMC Fit to Synthetic Afterglow Data}\label{sec:MCMC_synthetic_data}
\begin{figure*}
    \centering
    \begin{minipage}[b]{1.\linewidth}
        \centering
        \includegraphics[width=0.9\linewidth]{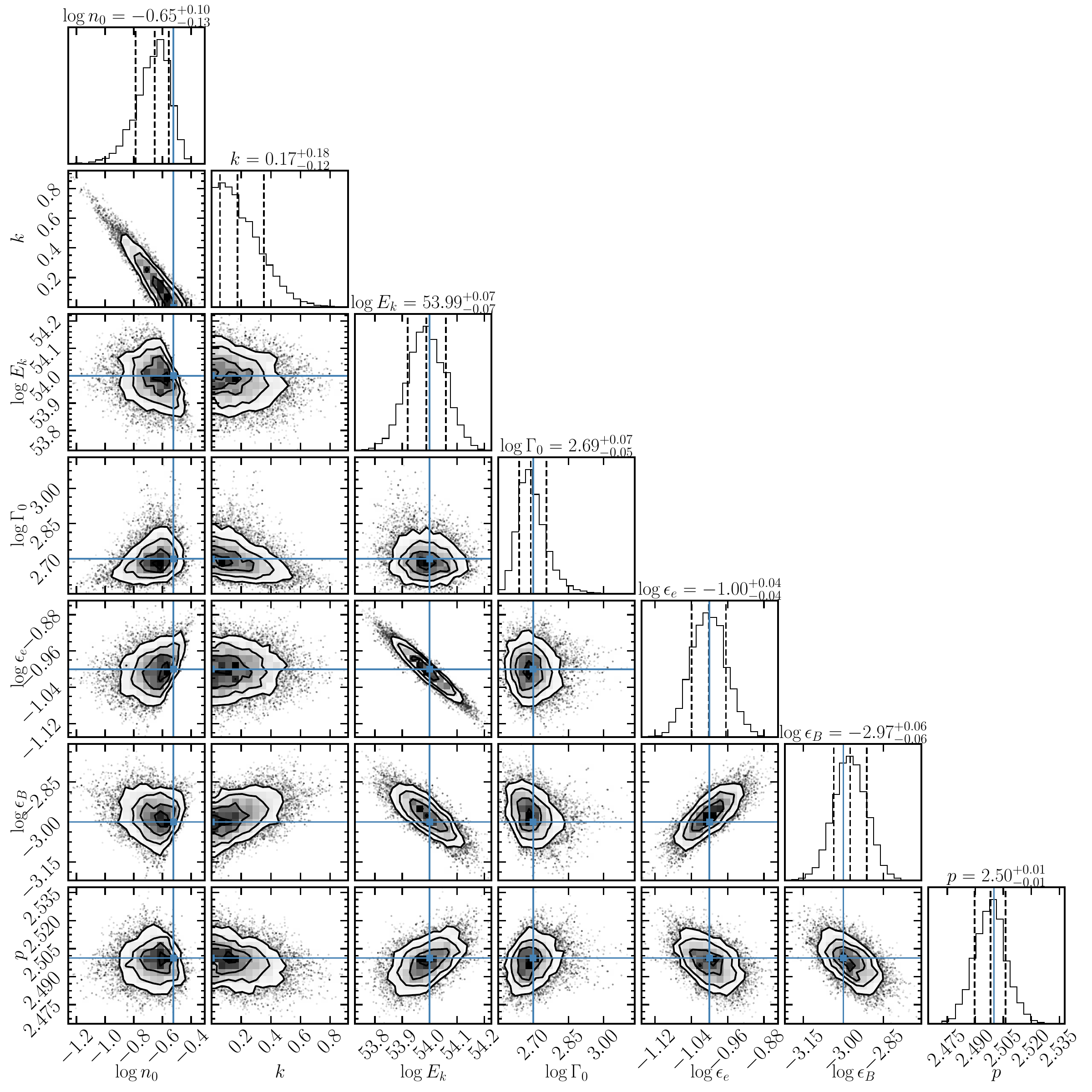}
    \end{minipage}\hfill 
    \begin{minipage}[b]{1.\linewidth}
        \centering
        \includegraphics[width=0.5\linewidth]{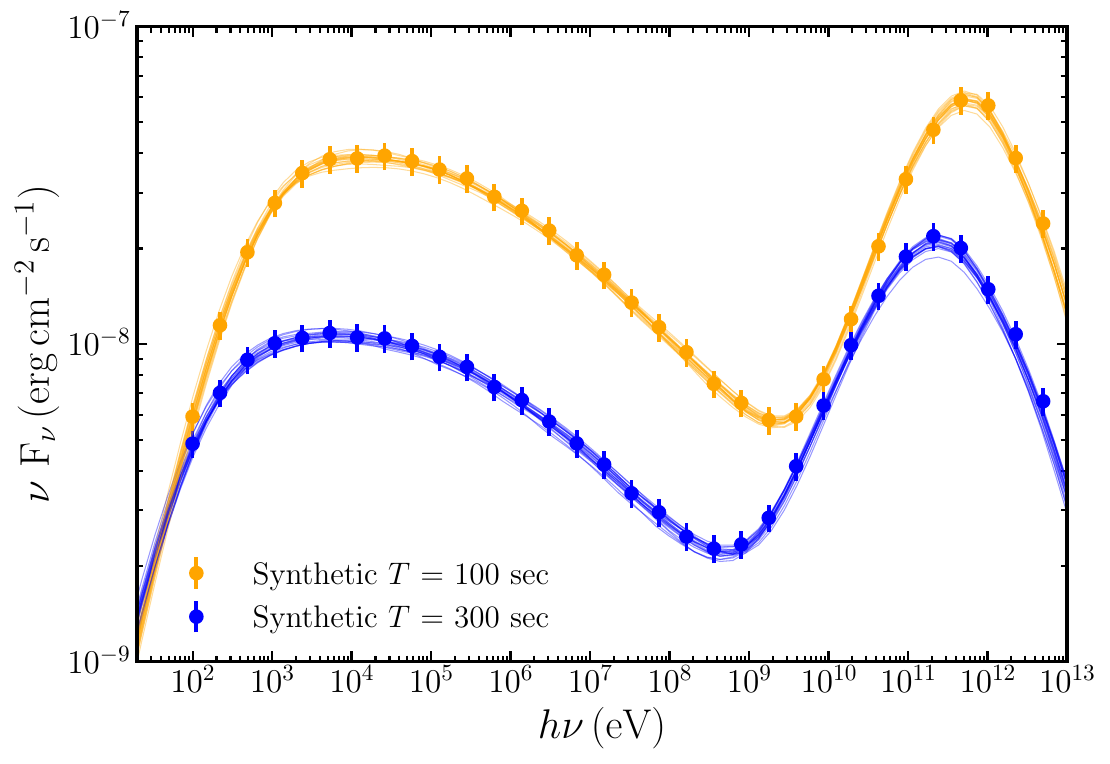}
    \end{minipage} 
   \vspace{-16pt} 
    \caption{ The upper panel show the posterior marginal probability distribution resulting of ran our MCMC sampler to fit synthetic data, blue crosses indicates the true values, and vertical lines indicates the median and the $1\sigma$ confidence level. The bottom panel show the synthetic data create for times $T=100 \, \rm sec$ and $T=200 \, \rm sec$ for the following parameters %
$E_{k, \rm iso} = 1 \times 10^{54} \, \rm erg$, $\Gamma_0 = 500$, 
$n_0 = 0.3 \, \rm cm^{-3}$, $k=0$, 
$\epsilon_e=10^{-1}$, $\epsilon_B=10^{-3}$, 
$p=2.5$, 
$z=0.43$ ($d_L = 2.45 \, \rm Gpc$) and the set of solutions within $1\sigma$ of confidence level.}
    \label{fig_CORNER_synthetic}
\end{figure*}

To test our MCMC sampler we perform a fit to synthetic spectra observed at two different times. We use the values assuming an early evolution of the afterglow spectra ( $T= 100 \, \rm sec$ and $T= 300 \, \rm sec$). The synthetic data was created taking 32 points covering the energy range between $10^2$ eV and $5 \times 10^{12}$ eV and assuming each point has an uncertainty equal to 10 per cent of the central value. Figure\,\ref{fig_CORNER_synthetic} shows the posterior distributions with the best-fit model parameters and the model spectra calculated by randomly sampling the posterior distributions. The five inferred parameter values from the MCMC are in excellent agreement within $1\sigma$ of confidence level around the true value. The remaining two correlated parameters, i.e.,  $n_0$ and $k$, are not so well constrained due to the sparse data that did not include the peak of the lightcurve, which would have constrained the deceleration radius. This limitation is common for GRB afterglows observations as early multi-band data are not available in geneal. Nevertheless, our best-fit values for these parameters are close to their true values. Therefore, we conclude that our MCMC sampler is reliable and capable of inferring the physical model parameters from broadband spectral data.

\end{document}